\documentclass[11pt,superscriptaddress,aps,showpacs,nofootinbib,longbibliography]{revtex4-1}

\pdfoutput=1

\usepackage{ulem}

\usepackage{comment}
\usepackage{hyperref}
\usepackage{graphicx}
\usepackage{amsmath,amssymb,color}
\usepackage{grffile}
\usepackage{ulem}
\usepackage{graphicx}
\usepackage{changepage}

\newcommand{\beq}{\begin{equation}}
\newcommand{\eeq}{\end{equation}}

\newcommand{\beqn}{\begin{eqnarray}}
\newcommand{\eeqn}{\end{eqnarray}}

\newcommand{\es}[1]{{\color{blue}[ES: #1]}}

\newcommand{\ud}{\,\mathrm{d}}

\newcommand{\Var}{\,\text{Var}}
\newcommand{\E}{\,\text{E}}

\begin{document}


\title{Robust non-minimal attractors in many-field inflation
}

\author{Perseas Christodoulidis}
\email{perseas@ewha.ac.kr}
\affiliation{Department of Science Education, Ewha Womans University, Seoul 03760, Republic of Korea}
\affiliation{Korea Institute for Advanced Study, Seoul 02455, Republic of Korea}

\author{Robert Rosati}
\email{robert.j.rosati@nasa.gov}
\affiliation{University of Alabama in Huntsville, Huntsville, AL 35812, USA}
\affiliation{NASA Marshall Space Flight Center, Huntsville, AL 35812, USA}

\author{Evangelos I.~Sfakianakis}
\email{evangelos.sfakianakis@austin.utexas.edu}
\affiliation{Department of Physics, Case Western Reserve University, Cleveland,
OH 44106, USA}
\affiliation{Texas Center for Cosmology and Astroparticle Physics,
Weinberg Institute for Theoretical Physics, Department of Physics,
The University of Texas at Austin, Austin, TX 78712, USA}
\affiliation{Department of Physics, Harvard University, Cambridge, MA, 02131, USA}

\begin{abstract}
{
Multi-field inflation can be inherently non-predictive, with the exception of models with strong attractors.
In this work, we focus on models with multiple scalar fields that are non-minimally coupled to the space-time Ricci curvature scalar, motivated by the expectation of a rich particle spectrum at high energies. 
%
We show that in this family of models, the single- and multi-field predictions for CMB observables are identical, as long as there exists at least one non-minimal coupling with $\xi \gg 1$.  We provide simple expressions for the Hubble scale, the number of $e$-folds, and the turn rate for systems with an arbitrary number of fields and explore the statistical properties for different priors. 
\\
}
\end{abstract}

\maketitle

\newpage 

\tableofcontents

\newpage

\section{Introduction}

Over four decades after being proposed, inflation remains the leading framework for pre-Big Bang cosmology \cite{Guth:1980zm, Linde:1983gd, Guth:2005zr}. In addition to its simplicity and elegance, inflation has been confronted with a series of experiments, space-, earth- and balloon-based. So far,  the inflationary paradigm has withstanded experimental scrutiny, despite several simple models being ruled out and others being in tension with observations \cite{Planck:2018jri}. 

Due to the success of the inflationary paradigm, a plethora of inflationary models have been proposed \cite{Martin:2013tda}. The majority of inflationary models can be described by (effectively) single-field dynamics, where the slow-roll attractor is reached and makes the predictions for the CMB observables independent of the initial conditions (within certain bounds). 
Despite the appeal of single-field models due to their simplicity, it is reasonable to expect the presence of many fields at high energies, which may be relevant for inflation. There is a rich literature on multi-field models, including the distinction between slow-roll and slow-turn conditions (see, e.g., Refs.~\cite{Cremonini:2010ua, Achucarro:2012yr}), as well as the ability of multi-field systems to support fast-turn trajectories \cite{Christodoulidis:2018qdw, Bjorkmo:2019fls, Garcia-Saenz:2018ifx, Achucarro:2019pux,Aragam:2020uqi,Anguelova:2022foz,Christodoulidis:2022vww,Anguelova:2024akm,Wolters:2024vzk}, leading to interesting phenomenology.

Going beyond models with two (or a few) fields, the dynamics can be both very complicated and highly dependent on initial conditions. This leads these models to produce a ``band of observables'', rather than a clear prediction for the CMB. 
This area has received less attention than single- and few-field systems, but important work has been done, specifically for sum-separable potentials \cite{Frazer:2013zoa} and quadratic $N$-flation \cite{Easther:2013rva}, using random matrix theory methods \cite{Dias:2016slx, Dias:2017gva,Paban:2018ole} or focusing on random Gaussian landscapes~\cite{Masoumi:2016eag, Masoumi:2017gmh, Masoumi:2017xbe}. Interestingly, in the case of quadratic $N$-flation, a peak distribution was shown to arise for the spectral tilt $n_s$, with a shift from the corresponding single-field value \cite{Easther:2013rva}. However,  this result was later shown to depend on the choice of initial conditions \cite{Christodoulidis:2019hhq}.

In this work, we focus on a family of multifield models with
non-minimal couplings to gravity of the form $\xi \phi^2 R$, where $\xi$ is the dimensionless non-minimal coupling of the scalar field $\phi$ to the space-time Ricci scalar $R$.
We consider this class of models to be well motivated for several reasons. Realistic models of particle physics include multiple scalar fields at high energies. In any such model,
non-minimal couplings are required for self-consistency, since they arise as renormalization
counterterms when quantizing scalar fields in curved spacetime \cite{Bunch:1980bs, Bunch:1980br, Callan:1970ze}. Moreover, the non-minimal coupling constants generically rise with energy under renormalization-group flow with no UV fixed point \cite{buchbinder1992}, and hence one expects $|\xi|\gg 1$ at inflationary energy scales. 

This class of models includes the well-known example of Higgs inflation ~\cite{Bezrukov:2007ep}, where the Standard Model Higgs plays the role of the inflaton, typically requiring a large non-minimal coupling $\xi \gtrsim{\cal O}(1000)$. Higgs inflation lends itself to a detailed analysis of preheating, as all couplings involved appear in the Standard Model \cite{Sfakianakis:2018lzf, Bezrukov:2008ut, Garcia-Bellido:2008ycs}. Furthermore, it has received numerous extensions (see, e.g. Refs.~\cite{He:2018gyf,Giudice:2010ka, Lerner:2010mq}).
Beyond the special case of Higgs inflation, non-minimally coupled
models have been extensively studied in the single- and two- (and three-) field cases, both during inflation \cite{Kaiser:2012ak, Kaiser:2013sna, Schutz:2013fua, Greenwood:2012aj}, as well as during preheating \cite{DeCross:2015uza,DeCross:2016fdz,DeCross:2016cbs,Nguyen:2019kbm,vandeVis:2020qcp, Ema:2016dny}. In particular, two- and three-field models with non-minimal couplings have been shown to possess a strong single-field attractor \cite{Kaiser:2013sna}, which persists during preheating \cite{vandeVis:2020qcp} and can be circumvented if potential parameters and initial conditions are concurrently fine-tuned \cite{Schutz:2013fua, Kaiser:2012ak}, leading to the generation of isocurvature modes and primordial non-Gaussianity.

We go beyond previous work on models with non-minimal couplings by studying them in the multi- and many-field regime. 
We derive analytical results for the attractor behavior of these models, going beyond the known two-field results~\cite{Kaiser:2012ak}, and show that they apply to systems with an arbitrary number of fields.
 This attractor allows this class of models to remain predictive, regardless of the number of fields involved in the inflationary process for a wide range of values for the potential parameters and initial conditions.
 We also verify these analytic results with a suite of numerical simulations up to $\mathcal{N} = 30$ non-minimally coupled fields, leveraging the high-$\mathcal{N}$ capabilities of \texttt{Inflation.jl} \cite{rosati_2020_4708348}.

The paper is organized as follows. In Section~\ref{sec:model} we describe the model and briefly list the known results derived for two fields. In Section~\ref{sec:attractor} we describe the attractor solutions that arise for an arbitrary number of fields and provide analytical results for the attractor which were missing from the literature, including formulas for the Huble scale and the turn rate.
We verify our analytical construction with numerical simulations. Section~\ref{sec:simulations} contains the results of our extensive numerical simulations for a variety of potential parameters, initial conditions, and number of fields, which demonstrate the emergence of a strong single-field attractor during inflation.
Distributions of important quantities, including the Hubble scale and the turn rate, are presented and analyzed. Despite the distribution of background parameters, the existence of the strong attractor leads to results for the adiabatic spectral tilt $n_s$, which are indistinguishable from the single-field result for all our simulations. 
We conclude and provide suggestions for future work in Section~\ref{sec:summary}.

\section{Model and formalism}
\label{sec:model}

We consider inflationary models consisting of multiple real scalar fields $\phi^I$, each non-minimally
coupled to the Ricci spacetime curvature scalar. We work in $(3 + 1) $ spacetime dimensions and adopt the ``mostly plus''
spacetime metric signature $(-,+, +, +)$. The action in the Jordan frame, where
the fields' nonminimal couplings remain explicit, is written as
\beq
\tilde S = \int \ud^4 x\sqrt{-\tilde g} \left [
f\left (\phi^I\right ) \tilde R
- \frac{1}{2}\tilde {\cal G}_{IJ} \tilde g^{\mu\nu} \partial_\mu \phi^I \partial_\nu\phi^J
-\tilde V\left (\phi^I\right )
\right 
] \, ,
\eeq
where $\tilde R$ is the spacetime Ricci scalar, $f\left (\phi^I\right )$ is the nonminimal coupling function, and $\tilde {\cal G}_{IJ} $ is the Jordan-frame
field space metric. For the remainder of this work, we choose $\tilde {\cal G}_{IJ}=\delta_{IJ}$, which can be thought of as the simplest choice, since it leads to canonical kinetic terms in the Jordan frame. We take the
Jordan-frame potential, $\tilde V\left (\phi^I\right )$, to have a generic, renormalizable polynomial form, with quartic self-interaction and two-to-two cross-couplings
\beq
\tilde V(\phi^I) = \frac{1}{4} \sum_{I=1}^N {\lambda_I  } (\phi^I)^4
+ \frac{1}{2} \sum_{I=1}^N \sum_{J=1}^{I-1} g_{IJ } (\phi^I)^2(\phi^J)^2
\, ,
\eeq
where we kept the sum explicit for clarity. We neglect any bare mass terms, in order to reduce the number of parameters. However, any mass that is significantly smaller than the Hubble scale during inflation would not affect the dynamics, so in order to get significant effects from bare masses, some amount of parameter tuning would be needed. That being said, the inclusion of bare mass terms (or any other potential term) is straightforward, using the 
formalism of Ref.~\cite{Kaiser:2012ak} and the techniques developed in the present work.

In the Einstein frame, where the gravitational part acquires the usual Einstein-Hilbert form, the action becomes
\beq
\label{eq:action}
S = \int \sqrt{-g}\,  \ud^4x \left [
{1\over 2} M_{\rm Pl}^2 R + {1\over 2} g^{\mu\nu} {\cal G}_{IJ} \partial_\mu \phi^I \partial_\nu \phi^J
+
V(\phi^I)
\right ] \, .
\eeq
While this can encompass any non-minimal coupling function $f(\phi^I)$, we choose the simple form (motivated by RG arguments \cite{buchbinder1992})
\beq
f(\phi^I) ={1\over 2} \left [ M_{\rm pl}^2 + \xi_I (\phi^I)^2 \right ]
\, ,
\eeq
where the non-minimal coupling parameters $\xi_I$ are in general different from each other.
In the Einstein frame, the fields are necessarily kinetically coupled with the  kinetic mixing matrix (equivalently, the field-space metric) 
\beq
{\cal G}_{IJ}(\phi^K) = {M_{\rm Pl}^2 \over 2f(\phi^I)} \left [  {\delta }_{IJ} +{3\over f(\phi^I)} f_{,I} f_{,J} \right ] \, , 
\label{eq:GIJ}
\eeq
where 
$
f_{,I} \equiv {\partial f\over \partial \phi^I} = \xi_I^2 \phi^I 
$
and no sum on $I$ is implied in the definition of $f_{,I}$.

The potential in the Einstein frame is  related to the potential in the Jordan frame through a conformal ``stretching''
\beq
V(\phi^I) = {M_{\rm Pl}^2 \over 4}{\tilde V(\phi^I)\over  f^2(\phi^I)} 
\, .
\eeq
We see that at large field values, the Einstein-frame potential develops smooth, asymptotically flat, directions, where slow-roll inflation with small tensor-to-scalar ration $r$ is realized.


We provide a brief derivation of the main equations for completeness. 
By varying the Einstein-frame action of Eq.~\ref{eq:action} with respect to the fields $\phi^I$, we arrive at the equations of motion
\beq
\Box \phi^I + \Gamma^I_{JK} \partial_\mu \phi^J \partial^\mu \phi^K - {\cal G}^{IJ}V_{,K}=0 \, ,
\eeq
where $\Box \phi^I\equiv g^{\mu\nu} \phi^I_{;\mu;\nu}$ and $ \Gamma^I_{JK} $ is the field-space Christoffel symbol, derived from the field space metric ${\cal G}_{IJ}$ of Eq.~\eqref{eq:GIJ}. By decomposing the field $\phi_I = \varphi^I + \delta\phi^I$ into the background field $ \varphi^I(t)$ and the fluctuations $\delta\phi^I(x^\mu)$, the background equations of motion become
\beq
{\cal D}_t \dot\varphi^I + 3H\dot\varphi^I + {\cal G}^{IJ} V_{,J}=0 \, ,
\eeq
where we define the field-space covariant derivative ${\cal D}_J A^I \equiv \partial_JA^I +\Gamma^I_{JK}A^K$ for a vector $A^I$. The covariant time derivative is defined as ${\cal D}_tA^I\equiv \dot\varphi^J {\cal D}_J A^I$ and $H=\dot a / a$ is the Hubble parameter,
which at background order obeys
\beqn
H^2 &=& {1\over 3 M_{\rm {Pl}}^2} \left [
{1\over 2} {\cal G}_{IJ} \dot\varphi^I \dot\varphi^J +V(\varphi^I)
\right ] \, ,
\\
\dot H &=& -{1\over 2 M_{\rm {Pl}}^2}{\cal G}_{IJ} \dot\varphi^I \dot\varphi^J \, .
\eeqn
We also define the Mukhanov-Sasaki variable  as
\beq
Q^I= {\cal Q}^I+ {{\dot\varphi}^I\over H}\psi \, ,
\eeq
where ${\cal Q}^I$ is a covariant fluctuation vector that reduces to $\delta\phi^I$ to first order in the fluctuations \cite{Gong:2011uw} and $\psi$ is the diagonal spatial perturbation to the spatially flat FRW metric $ds^2 \subset a^2(t) (1-2\psi) \delta_{ij} dx^i dx^j$.
The equations for the Mukhanov-Sasaki variables can be similarly written in a covariant form \cite{Sasaki:1995aw}
\beq
{\cal D}_t^2 Q^I +3H {\cal D}_t Q^I + 
\left [
{k^2\over a^2}\delta^I_J + {\cal M}^I_{\, J}
\right ]Q^J=0 \, .
\eeq
The mass matrix ${\cal M}^I_{\, J}$ is
\beq
{\cal M}^I_{\, J}\equiv{\cal G}^{IK} {\cal D}_J{\cal D}_K V - {\cal R}^I_{LMJ}\dot\varphi^M - {1\over a^3 M_{\rm {Pl}}^2} {\cal D}_t\left (
{a^3\over H} \dot\varphi^I \dot \varphi_J 
\right ) \, ,
\eeq
where ${\cal R}^I_{LMJ}$ is the field-space Riemann tensor. We can safely neglect the last term in the above equation, since it is  slow-roll suppressed during inflation. 

As usual, we will study perturbations in the kinematic frame, an orthonormal basis in  field space \cite{Gordon:2000hv}. We choose the first vector as the unit tangent vector of the field-space trajectory
\beq
  \sigma^I \equiv {\dot \phi^I\over \dot\sigma} \, ,
\eeq
where $\dot\sigma$ denotes the magnitude of the velocity of the background motion
\beq
\dot \sigma \equiv |\dot\phi^I| = \sqrt{{\cal G}_{IJ} \dot\phi^I\dot\phi^J}\, .
\eeq
Taking the covariant time derivative of this allows us to define the normal vector through
\beq
{\cal D}_t \sigma^I \equiv \omega   n^I \, ,
\eeq
where $\omega$ is the turn rate which quantifies the (local) shape of the trajectory and how it deviates from a straight line. The covariant derivative of the normal vector defines the binormal vector
\beq
{\cal D}_t  n^I \equiv \tau   b^I -\omega   \sigma^I \, ,
\label{eq:binormal}
\eeq
where $\tau$ is the torsion, quantifying how the trajectory twists away from a plane. In a similar fashion, we can define higher torsion parameters $\tau_i$ that measure the rate of change of the basis vectors $\{E^I_\alpha\}$. They are calculated explicitly from the relation 
\beq \label{eq:rotation matrix}
{\cal{D}}_t E^I_{\alpha} = z^{~\beta}_{\alpha} E^I_\beta \, ,
\eeq
where $z$ is an antisymmetric tensor with non-zero elements only on the diagonals above and below the principal one.

In terms of the curvature perturbation ${\cal R}$ and $\mathcal{N}-1$ normalized isocurvature perturbations defined through
\beq
\mathcal{R} \equiv \frac{Q^I \sigma_I}{\sqrt{2\epsilon}} \, , \quad Q_{\rm n} \equiv Q^I   n_I \, , \quad Q_{\rm b} \equiv Q^I   b_I \, , \quad\dots
\eeq
the corresponding equations of motion are (see e.g. \cite{Achucarro:2018ngj})
\begin{align}
& {\ud \over \ud t} \left( \dot{\mathcal{R}} - 2 {\omega \over \sqrt{2\epsilon}}  Q_{\rm n}\right) + (3 + \eta) H \left( \dot{\mathcal{R}} - 2 {\omega \over \sqrt{2\epsilon}}  Q_{\rm n}\right) =  {\nabla^2 \mathcal{R} \over a^2}  \, , \\ \nonumber
& {\ud \over \ud t} \left(\dot{Q}_{\alpha} -  z^{~\beta}_{\alpha} Q_\beta \right) + 3H \left(\dot{Q}_\alpha -  z^{~\beta}_{\alpha} Q_\beta \right) =  - 2  \sqrt{2\epsilon} \omega  \left( \dot{\mathcal{R}} - 2 {\omega \over \sqrt{2\epsilon}}  Q_{\rm n}\right) \delta_{\rm n\alpha} \\
& \qquad + z^{~\beta}_{\alpha} \left(\dot{Q}_\beta -  z^{~\gamma}_{\beta} Q_\gamma\right)  +  {\nabla^2 Q_\alpha  \over a^2 }     - M_{\alpha \beta} Q_{\beta} - 3 \omega^2 \delta_{\rm n\alpha} Q_{\rm n} \, ,
\end{align}
where $M_{\alpha \beta} \equiv E^I_\alpha E^J_\beta {\cal M}_{IJ}$ and $\alpha,\beta=1,\dots,\mathcal{N}-1$
denote the orthogonal fields. For the rest of this work, we will consider slow-turn evolution and thus ignore terms proportional to $\omega^2$ and the rotation matrix $z^{~\beta}_{\alpha}$, as defined in Eq.~\eqref{eq:rotation matrix}.
\clearpage

\section{Attractor behavior}
\label{sec:attractor}

\subsection{Background dynamics}

It has been well established that two-field models with non-minimal couplings possess strong single-field attractors at strong coupling $\xi\gg1$ \cite{Kaiser:2012ak, Kaiser:2013sna, Schutz:2013fua, DeCross:2015uza}. In simple cases, where the attractor is aligned with one of the fields, leading to motion only along $\phi^J$, the attractor solution is trivial ($\phi^I=0$ with $I\ne J$) and the potential along this attractor is (to lowest order in $1/\phi_J$)
$V \simeq {\lambda_J}M_{\rm {Pl}}^4/{(4\xi_J^2)}$.
However, this is only true for motion along one of the axes in field space. 
Taking the two field case as an example,  the attractor in the general case will be along some angle in the $\phi^1-\phi^2$ plane, defined by a combination of the potential parameters and non-minimal couplings \cite{Schutz:2013fua}. In this Section, we go beyond the known two-field results,  deriving the properties of the attractor for systems with an arbitrary number of fields.
 
The attractor behavior becomes manifest in a coordinate system where one field drives inflation and all other fields are frozen at the minima of their effective potential \cite{Christodoulidis:2019mkj}. In the case of zero cross-couplings $g_{IJ}=0$ 
the Einstein-frame potential has the simple form
\beq
 V = { M_{\rm {Pl}}^4 \over 4} {\sum_{I=1}^{\mathcal{N}}\lambda_I (\phi^I)^4
 \over \left (1+ \sum_{I=1}^{\mathcal{N}} \xi_I (\phi^I)^2 \right )^2} \, .
\eeq
We perform the substitution 
\begin{align}
r^2  &= {\sum_I \xi_I (\phi^I)^2} \, , \\ \label{eq:xi}
x^I &= \frac{1}{r}\sqrt{\xi_I}  \phi^I   \, ,
\end{align} 
with no sum on $I$ implied in Eq.~\eqref{eq:xi}, for which the potential simplifies to a product-separable form
\beq
V = { M_{\rm {Pl}}^4 \over 4} {r^4 \over (1+r^2)^2} \sum_I {\lambda_I \over \xi_I^2} (x^I)^4  \, .
\eeq
In this parametetrization the $x^I$ coordinates are constrained on the unit hypersphere in the $\mathcal{N}$-dimensional space and we can express the last of them in terms of the other $\mathcal{N}-1$ 
\beq
\left (x^{\mathcal{N}}\right)^2 =  {1 -  \sum_i^{\mathcal{N}-1} \left (x^i\right )^2 } \, ,
\eeq
where lowercase Latin indices take values $i=1,2,\dots,\mathcal{N}-1$. We call the metric in this coordinate system $h_{IJ}$, which can be written as
\begin{align} \label{eq:hrr}
h_{rr} &= {1 \over 1+r^2} \sum_K^{\mathcal{N}} {(x^K)^2 \over \xi_K}  + {6 r^2 \over (1+r^2)^2} \, , \\ \label{eq:h_ra}
h_{r i} &=  {x^{i} r \over 1 + r^2} \left( {1 \over \xi_{i} } - {1 \over \xi_{\mathcal{N} }} \right) \, ,\\ \label{eq:hab}
h_{ij} &=  {r^2 \over 1 + r^2} \left( {\delta_{ij} \over \xi_{a} } + { x^{i} x^{j} \over \xi_{\mathcal{N} } \left(x^{\mathcal{N} }\right)^2} \right) \, ,
\end{align}
where 
no sum on $i$ is implied in Eq.~\eqref{eq:h_ra}. Looking at the components of the metric, we observe that at large values of $r$ all components except for $h_{ij}$ scale as $r^{-2}$. Moreover, the second term in Eq.~\eqref{eq:hab} is strictly smaller than the first because $|x^{i}|<1$ and we have assumed (without loss of generality) that $\xi_{\mathcal{N}}>\xi_{i}$. Therefore, in what follows, we can treat the metric as almost diagonal for $r \gg 1$. 

On the attractor solution, the fields $x^i_{\rm at}$ are given as solutions of the following set of equations
\beq \label{eq:effective_gradient}
V^{,i}_{~\rm eff} \equiv \Gamma^i_{rr} (r')^2 H^2 +V^{,i} \simeq 0 \, .
\eeq
At large $r$ and up to $\mathcal{O}(r^{-2})$ corrections the solution of the previous equations coincides with the solution of $V_{,i}=0$, i.e.~we find that the orthogonal fields $x^i$ are stabilized at values close to the minima of $V$ in the $x$ direction.\footnote{This is a highly non-trivial result which deserves further elaboration. The minima of the potential in the $x$ direction, $V_{,i}(x_{,i}^{\rm min})=0$, are not minima of the effective gradient since the Christoffel symbols do not vanish at these points. Denoting the solutions to $V_{\rm eff}^{,i}=0$ as $\tilde{x}^{\rm eff}_i$ and expanding around them as $\tilde{x}^{\rm eff}_i=x_{i}^{\rm min}+\delta x_i$, to first order in $\delta x_i$ we find that the gradient term satisfies $V^{,i}\approx h^{ij} V_{,j}\approx \xi_{i}V_{,ii} \delta x_i$, which at strong coupling can be $\mathcal{O}(1)$. This term is canceled to leading order in slow roll by the Christoffel term, leaving overall $V^{,i}_{\rm eff}=\mathcal{O}(\epsilon)$; see also Ref.~\cite{Schutz:2013fua}.} 
These minima satisfy
\beq \label{eq:x_k}
x^i_{\rm at} = \pm {\xi_{i}/ \sqrt{\lambda_i} \over \sqrt{\sum_{K=1}^{\mathcal{N}}  \xi_{K}^2}/\lambda_I} \, ,
\eeq
leading to the potential along the attractor having the closed form solution
\beq \label{eq:vattractor}
V_{\rm at} = \frac{ r^4}{4(1+r^2)^2} \frac{M_{\rm {Pl}}^4}{\sum_{I=1}^{\mathcal{N}} \xi_I^2/\lambda_I }\, .
\eeq
This surprisingly simple and accurate formula has not appeared in the literature until now.
The corresponding Hubble scale is given by $H^2 \simeq V/(3M_{\rm Pl}^2)$ during slow-roll inflation. Under the same approximation we find that the $r$ field follows its potential gradient
\beq \label{eq:dr}
r' \approx -V^{,r} \approx -h^{rr} (\log V)_{,r}
\eeq
and that the first slow-roll parameter receives its dominant contribution from the $r$ field
\beq
\epsilon \approx {1 \over 2} h^{rr} (\log V)_{,r} \, .
\eeq
Using Eq.~\eqref{eq:dr} we find that the number of $e$-folds  as a function of the radius is given by
\beq \label{eq:Nef_approx}
\Delta N \approx - \int {V \over h^{rr} V_{,r}} \ud r  \approx  {3 \over 4} ( r^2 -r_{\rm end}^2 )+  \mathcal{O}({\xi_{\mathcal{N}}}^{-1} )\, ,
\eeq
where $r_{\rm end}$ refers to the end of inflation. The case of 
non-zero cross-couplings $g_{IJ}$ is intuitively straightforward, even if algebraically more cumbersome in the case of many fields. Appendix~\ref{app:attractor} contains a detailed derivation of the attractor in this most general case.  

\begin{figure}
\includegraphics[width=.45\textwidth]{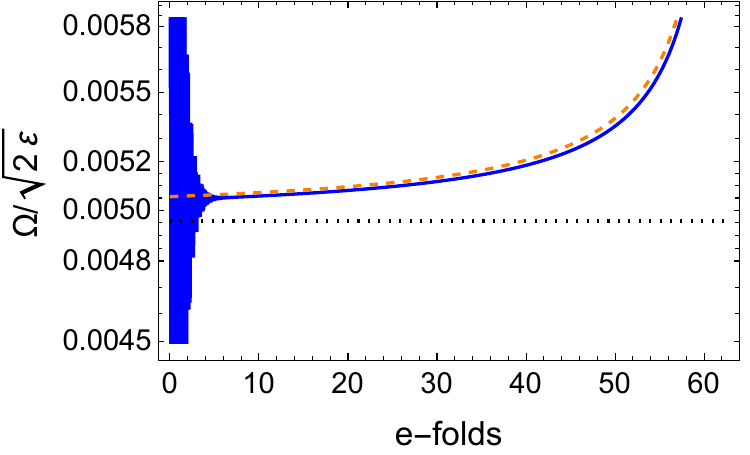}
\caption{
The normalized turn rate divided by $\sqrt{2\epsilon}$ for a three-field realization with $\lambda_1=1,~\lambda_2=2,~\lambda_3=2,~\xi_1=10,~\xi_2=90,~\xi_3=200$. The plot depicts a total of  61 $e$-folds  of inflation. The blue solid line shows the full numerical result, where the initial large oscillations represent the transient behavior until the system reaches the attractor.
The orange-dashed curve shows the  analytical approximation of Eqs.~\eqref{eq:Nef_approx} and~\eqref{eq:turn_rate}.
The asymptotic value of Eq.~\eqref{eq:asymptotic2} is shown by the horizontal black-dotted line.  }
\label{fig:turn_app}
\end{figure}

The normalized turn  rate is given by \footnote{The normalized turn rate in general can be calculated through
$
\Omega^2  \equiv \frac{\omega^2}{H^2} = {1 \over 2\epsilon } G_{IJ} \, D_N (\phi^I)' D_N (\phi^J)' - {1 \over 4}\eta^2 $.
} 
\beq \label{eq:turn_rate}
\Omega^2\equiv \frac{\omega^2}{H^2}  \approx 2 \epsilon \, (h_{rr})^{-2}  h_{a b}  \Gamma^a_{rr}\Gamma^b_{rr} \approx 2\epsilon \sum_a {1 \over 4 h_{a a}} \left[(\log h_{rr})_{,a}\right]^2 \, ,
\eeq
where in the last step we used the diagonal metric approximation for $h_{IJ}$. The excellent accuracy of our analytical approximation is shown in Figure~\ref{fig:turn_app}.
We can also extract the asymptotic value for $r\gg 1$, deep into inflation. For two fields this is explicitly
\beq  \label{eq:asymptotic}
\Omega^2 \approx 2\epsilon { \left(1-R_{\rm \xi} \right)^2 \left( R_{\lambda}  +R_{\rm \xi}^3 \right) R_{\lambda} \, R_{\rm \xi} \, \xi \over \left( R_{\lambda} \, + R_{\rm \xi}^2 \right) \left[ R_{\lambda} +R_{\rm \xi}^2 + 6 \left( R_{\lambda} +R_{\rm \xi}^3 \right) \xi \right]^2} 
\simeq
{2\epsilon\over \xi} { \left(1-R_{\rm \xi} \right)^2 R_{\lambda} \, R_{\rm \xi}  \over 36 \left( R_{\lambda} \, + R_{\rm \xi}^2 \right)  \left( R_{\lambda} +R_{\rm \xi}^3 \right) } 
\, ,
\eeq
where the final expression holds in the limit of $\xi\gg 1$. We define
$R_{\rm \xi} \equiv \xi_1/\xi_2$ and $R_{\lambda} \equiv \lambda_1/\lambda_2$ (with $R_\xi,R_\lambda\le 1$), and we see that $\Omega^2$ is independent of $r$ up to $\mathcal{O}(r^{-2})$ which yields $\Omega \propto \sqrt{\epsilon / \xi}$ during inflation for $\xi \gg 1$. Similarly, for three fields the asymptotic value is
\beq 
\begin{aligned} \label{eq:asymptotic2}
\Omega^2 \approx & 2\epsilon {  \left(R_{\lambda_2}  R_{\rm \xi_1}^3 + R_{\lambda_1}  R_{\rm \xi_2}^3 + R_{\lambda_1} R_{\lambda_2} \right) R_{\lambda_1} R_{\lambda_2}  \, \xi 
\over \left( R_{\lambda_2} R_{\rm \xi_1}^2 + R_{\lambda_1} R_{\rm \xi_2}^2 + R_{\lambda_2}  R_{\lambda_1} \right)}  \times \\
& { R_{\lambda_2} R_{\rm \xi_1}\left(1 - R_{\rm \xi_1} \right)^2 + R_{\lambda_1}R_{\rm \xi_2} \left(1 - R_{\rm \xi_2} \right)^2 + R_{\rm \xi_1} R_{\rm \xi_2} \left( R_{\rm \xi_1} - R_{\rm \xi_2}\right)^2 \over  \left[ R_{\lambda_1} R_{\lambda_2} + R_{\lambda_2} R_{\rm \xi_1}^2 + R_{\lambda_1} R_{\rm \xi_2}^2 + 6 \left( R_{\lambda_1} R_{\lambda_2} + R_{\lambda_1} R_{\rm \xi_2}^3
 + R_{\lambda_2} R_{\rm \xi_1}^3 \right) \xi \right]^2}
\end{aligned}\, ,
\eeq
where $R_{\xi_i} \equiv \xi_i/\xi_{\mathcal{N}}$ and $R_{\lambda_i} \equiv \lambda_i/\lambda_{\mathcal{N}}$.
\begin{figure}
\includegraphics[width=.45\textwidth]{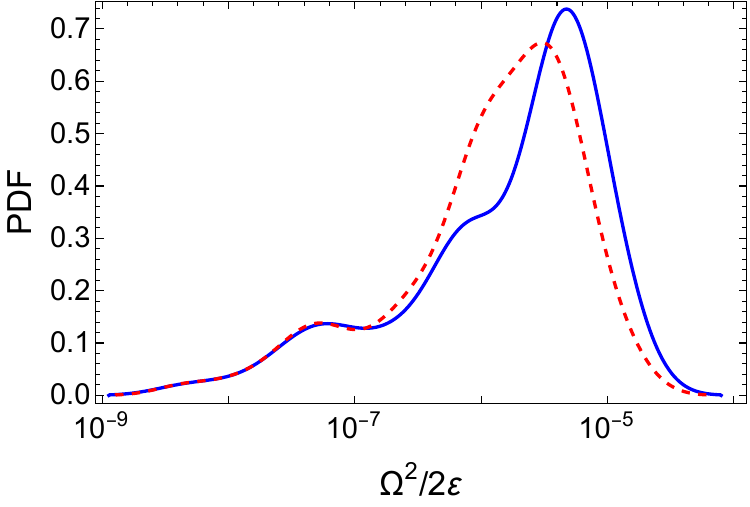}
\includegraphics[width=.45\textwidth]{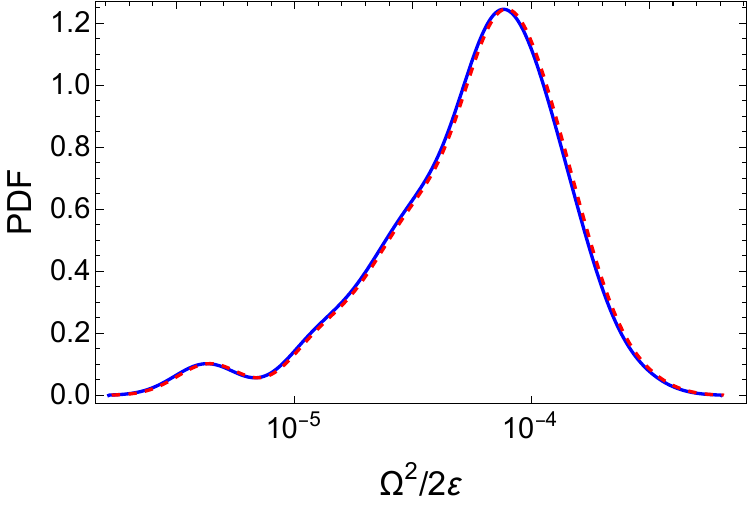}
\caption{The distribution of the normalized  turn rate  divided by the first slow roll parameter $\Omega^2/2\epsilon$, for two and three fields (left and right respectively) for 100 different realizations, each  leading to to approximate  $100$ $e$-folds of inflation on the attractor solution. The blue-solid curves show the numerical results calculated 75 $e$-folds before the end of inflation, whereas the red-dashed curves show the asymptotic analytical result from Eqs.~\eqref{eq:asymptotic} and \eqref{eq:asymptotic2} respectively.
 The sets of parameters $\{\xi_i\}$ and $\{\lambda_i\}$ are drawn from a uniform distribution ranging between 1-100 and 1-10 respectively.}
\label{fig:omega_test1}
\end{figure}
Figure~\ref{fig:omega_test1} shows the excellent agreement between our analytical expressions and the full numerical evolution of the system.

\begin{figure}
\includegraphics[width=.45\textwidth]{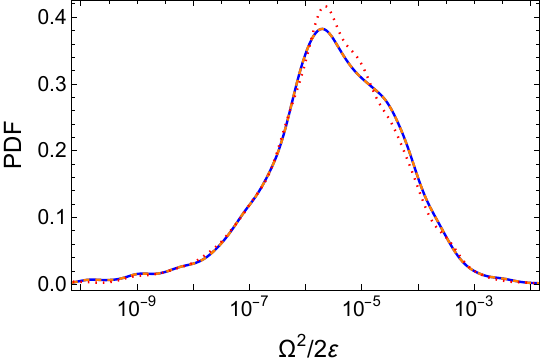}
\includegraphics[width=.45\textwidth]{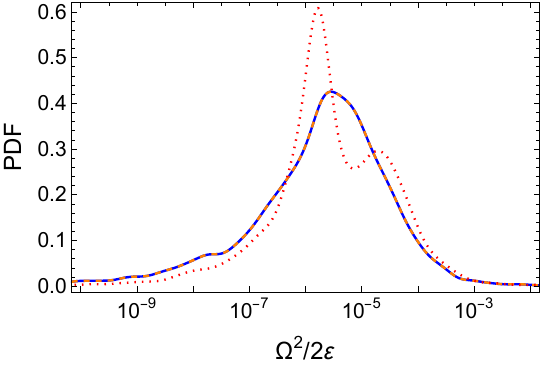}
\caption{The distribution of the  asymptotic turn rate normalized by the first slow roll parameter $  \Omega^2/2\epsilon$, as in Eq.~\eqref{eq:asymptotic}, for two fields and $10000$ different realizations. In both panels, $\xi$ is drawn from a uniform distribution with $\xi\in [75,125]$, whereas $\lambda$ is drawn from a uniform distribution (left) or a Marchenko-Pastur distribution (right). In both panels the distributions have the same averages values $\bar{\lambda}=$ $1$ (blue), $10$ (orange) and $100$ (red) and standard deviation $\sigma_\lambda = \bar{\lambda}/4$. The parameters of the Marcencho-Pastur are $\sigma_{\rm MP} = \left ( {\bar{\lambda}}\right )^{1/2}$ and $\beta_{\rm MP} = 1/2$.}
\label{fig:omega_Rscan1}
\end{figure}

It is instructive to explore how different distributions of parameters $\lambda$ and $\xi$ affect the distribution of the turn-rate. We see that, apart from the single appearance of $\xi$ in the denominator that provides the overall scaling $\Omega^2\propto 1/\xi$, the distributions of parameters enter into the calculation of the ratios $R_\lambda$ and $R_\xi$. Thus, the absolute magnitude of the self-coupling does not affect the turn rate (as expected, since $\lambda$ simply provides the overall scale of the potential). Drawing $\lambda_I$ from a uniform distribution with ever-changing spread, we see that the distribution of the turn-rate remains practically unchanged. This is due to a unique property of the uniform distribution: the distribution of the ratio is also uniform and since it is bounded between $0$ and $1$, the height of the distribution is also unity. Thus the spread of the distribution of $\lambda_I$ does not affect the distribution of $\Omega^2\xi$, as long as the former is uniform, as shown in the left panel of Fig.~\ref{fig:omega_Rscan1}.
That being said, doing the same numerical experiment, but drawing $\lambda$ from different realizations of the Marcenko-Pastur distribution, we see that the peak value of $\Omega^2/2\epsilon$ remains largely stable, but the shape can change, as seen on the right panel of Fig.~\ref{fig:omega_Rscan1}.
Before proceeding, we must stress a subtlety regarding the presentation of the probability density function of the turn rate. In order to avoid an extreme ``contamination" of our results by a single large value, which may appear in our simulation due to numerical errors, all PDFs of the turn rate are calculated as PDF's of the logarithm of the turn-rate and presented for visual simplicity as PDF's of the turn rate on a logarithmic scale.

\begin{figure}
\includegraphics[width=.45\textwidth]{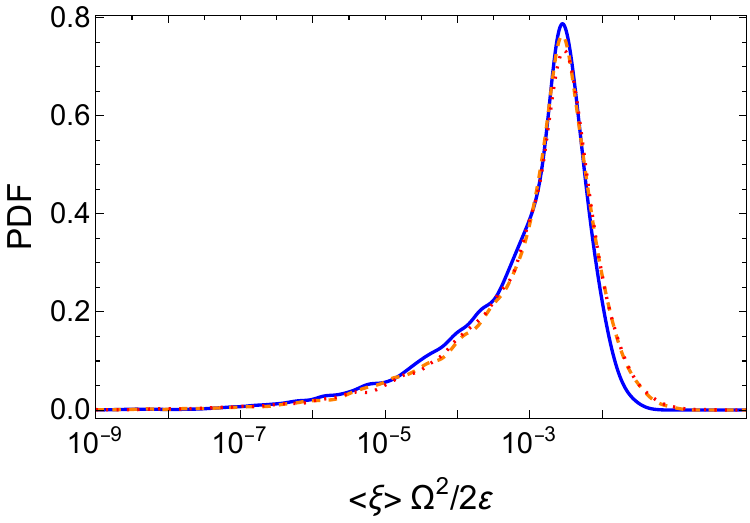}
\includegraphics[width=.45\textwidth]{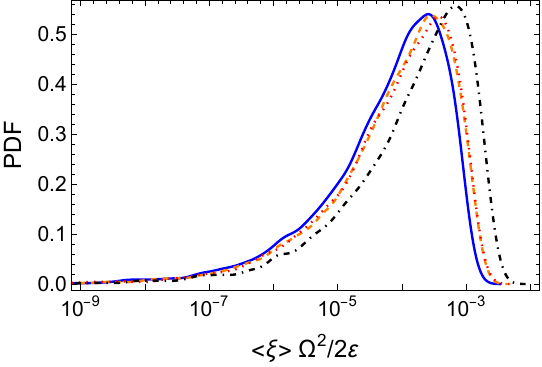}
\caption{ The distribution of the  asymptotic turn rate normalized by the first slow roll parameter and the average non-minimal coupling 
$\langle\xi\rangle \Omega^2/2\epsilon$ (right) 
for two fields and $10000$ different realizations. The self-couplings $\lambda$ are drawn from a uniform distribution with $\lambda\in[1,100]$. 
{\it Left:} The non-minimal couplings $\xi$ are drawn from a uniform distribution with values $\xi\in[1,10]$ (blue),
$\xi\in[1,100]$ (orange-dashed) and $\xi\in[1,1000]$ (red-dotted).
{\it Right:} The non-minimal couplings $\xi$ are drawn from a Marchenko-Pastur distribution with values $\bar \xi =1$ (blue),
$\bar \xi=10$ (orange-dashed) and $\bar \xi=100$ (red-dotted). The standard deviation for each curve is set as $\sigma_\xi=\bar\xi/4$. The black dot-dashed curve has $\bar \xi=100$ and $\sigma_\xi=\bar\xi/3$.
}
\label{fig:omega_Rscan2}
\end{figure}

We now move to examine the dependence on the turn rate on the distribution of the non-minimal couplings $\xi$.
Since for each set of parameters, the normalized turn rate is inversely proportional to the largest non-minimal coupling $\Omega/2\epsilon \propto 1/\xi$, we expect the distribution of $\Omega/2\epsilon$ to shift, depending on the ``typical" value of $\xi$. 
In this case, it is worth examining the rescaled distribution $\langle \xi\rangle \Omega^2/2\epsilon$.
The left panel of Fig.~\ref{fig:omega_Rscan2} shows that the rescaled turn rate is invariant with respect to the 
limits of the $\xi$ distribution, provided that it is uniform. This was somewhat expected, following our discussion on the dependence of the turn rate on the distribution of $\lambda$. Drawing  $\xi$ from a Marchenko-Pastur distribution 
still leads to an almost unchanged distribution of $\langle \xi\rangle \Omega^2/2\epsilon$, as long as the ratio of $\mu_\xi/\sigma_\xi$ is unchanged. If on the other hand, we do vary this ratio, the shape and peak of $\langle\xi\rangle \Omega^2/2\epsilon$ change, due to the large skewness of the Marchenko-Pastur distribution. Nevertheless, the overall scaling $\langle \Omega^2/2\epsilon\rangle \propto 1/\langle\xi\rangle$ appears to be rather universal for this type of model, with a prefactor that depends on the specific distributions that are chosen for the various parameters of the potential. We will return to this point when we examine the parameter dependence of all background quantities in Section~\ref{sec:Generalcase}.

\subsection{Perturbations}

Since the turn rate and the rest of the torsion parameters are small,\footnote{The geometrical structure of the attractor leads to an almost straight trajectory. This is verified by the small turn rate and hence all torsion parameters are expected to be even smaller.}  we can consider all perturbations to be uncorrelated at horizon crossing and be independently initialized at the Bunch-Davies vacuum. At superhorizon scales, the behavior of linear perturbations will be determined by the magnitude of the eigenvalues of the isocurvature mass matrix on these scales
\cite{Peterson:2011yt,Kaiser:2012ak,Schutz:2013fua,Christodoulidis:2022vww}. Approximating  the $x$-fields as ``frozen" in the minima of their respective effective potentials, the tangent vector points along the $r$ direction 
\beq
\sigma^I \approx (\sqrt{h^{rr}},0,\cdots) \, ,
\eeq
and the turning vector has non-zero components only along the orthogonal subspace 
\beq
\Omega n^I \approx  \sqrt{2\epsilon} \Gamma^I_{rr} (h_{rr})^{-1} \approx  \sqrt{2\epsilon} h^{II} (\log h_{rr})_{,I}  \, .
\eeq
Similarly, we obtain the mass matrix in the orthogonal subspace written in the kinematic basis as
\beq
\mathcal{M}_{a b} \approx E^I_{a} E^J_{ b} \left({V_{;IJ} \over H^2} - 2 \epsilon h^{rr} R_{r IJ r} \right)\, ,
\eeq
where we  neglected contributions from the turn rate and the rest torsion parameters since they are small and we used the fact that the tangent direction coincides with $r$.\footnote{Before proceeding to the calculation of the mass matrix, we should note a peculiar feature of the geometric term in the definition of the mass matrix. For two fields this term can always be written as $\epsilon R$, where $R$ is the Ricci scalar, and at the strong coupling limit $\xi \gg 1$ we find that $R$ approaches an $\mathcal{O}(1)$ constant which can be either positive or negative. Therefore, during inflation, this term is subdominant and the potential term dominates the effective isocurvature mass. However, for $\mathcal{N}>2$ the Ricci scalar becomes proportional to the non-minimal coupling for $\xi\gg 1$, i.e.~$R \sim \xi$ hence the space becomes highly curved. In contrast, the Riemann tensor components that enter the definition of the mass matrix remain $\mathcal{O}(1)$ and we find that the geometrical term becomes subdominant even though the field-space curvature could be very large and negative.} Furthermore, $h^{rr} R_{r IJ r}$ is found to be $\mathcal{O}(1)$ and thus the mass matrix is dominated by the potential term.

Having shown that contributions from geometrical terms in the mass matrix are subdominant, to determine the fate of perturbations on superhubble scales we need to examine the covariant derivatives of the potential
\beq
\mathcal{V}_{\alpha \beta} \approx 
E^I_{\alpha} E_J^{ \beta} {V_{; I}^{~J} \over H^2} \,.
\eeq
Since $ E^I_{A}$ is the inverse of $E_J^{ B}$, the eigenvalues of $\mathcal{V}_{\alpha \beta}$ coincide with the eigenvalues of the mixed Hessian $\mathcal{V}^I_{~J}$. In the latter, it can be shown that the Christoffel terms can be ignored, and thus we obtain
\beq
V_{; i}^{~j} \approx G^{jk} V_{,ik} \, .
\eeq
with $ G^{jK}\Gamma^L_{iK} V_{,L} \sim \mathcal{O}(\epsilon)$ and the eigenvalues of $V^{~j}_{;i}$ satisfying $\lambda_n \sim 24 \xi_n$. Thus, the isocurvature perturbations are strongly suppressed on superhubble scales, and the power spectrum freezes shortly after it crosses the horizon yielding predictions identical to the single-field predictions of non-minimal models.

It is worth comparing this type of slow-turn model with more commonly found models in the literature such as $\mathcal{N}$-flation. The latter models belong to the gradient flow class where every field follows its potential gradient
\beq
3H\dot{\phi}^I \approx - V^{,I} \, .
\eeq
In the absence of geometrical contributions from the Christoffel terms, the late-time attractor is reached when every field, except one, has become non-dynamical. Therefore, these fields should be placed at points that satisfy $V^{,I}=0$. If there is no hierarchy between these gradient terms,  this phase will be reached close to the end of inflation, leading to a family of possible field-space trajectories as the system approaches the global minimum of the potential. Due to the different possible ways the system can reach the minimum of the potential, predictions are not unique but rather depend on the initial field configuration. This is indeed the case for $\mathcal{N}$-flation with a range of masses and no strict hierarchy between them, where this behavior persists even at the many-field limit \cite{Christodoulidis:2019hhq}. In the case of non-minimal couplings, the existence of one or more  large couplings $\xi$  and otherwise generic  potential parameters is sufficient to quickly drive the system  to the attractor, practically in a few $e$-folds. It is worth noting, that in this context a ``large" non-minimal coupling needs to simply be larger than ${\cal O}(10)$.

\section{Many-field simulations}
\label{sec:simulations}

\subsection{Number of $e$-folds}
\label{sec:simbackground}

The first criterion for any successful model of inflation is its ability to sustain at least $60$ $e$-folds of inflation. In order to evaluate the dependence of the number of $e$-folds on the choice of priors for the parameters and initial conditions, we perform a number of simulations with different choices. 
The simulations were performed using \texttt{Inflation.jl} \cite{rosati_2020_4708348}, a Julia-language inflationary solver for both the background and the perturbations. \texttt{Inflation.jl} is based on the differential equations solvers in \texttt{DifferentialEquations.jl} \cite{rackauckas2017differentialequations}, the symbolic utilities of \texttt{SymPy.jl}\footnote{\url{https://github.com/JuliaPy/SymPy.jl}, a wrapper around SymPy \cite{sympy}. The model of this work pushed the limits of \texttt{SymPy.jl}'s symbolic capabilities, so we also generated alternative equations of motion using either automatic differentiation with \texttt{ForwardDiff.jl} \cite{RevelsLubinPapamarkou2016} or Mathematica expressions translated to Julia with \url{https://github.com/rjrosati/mathematica2julia}. These additional symbolic capabilities will be released in a future version of \texttt{Inflation.jl}.}, and the transport method \cite{Dias:2015rca,Dias:2016rjq}. Due to its use of common sub-expression elimination and modern differential equations methods, \texttt{Inflation.jl} can generate and solve the background and transport equations at high $\mathcal{N}$ for an arbitrary model of the form \eqref{eq:action}.

We always set the initial field velocities to  zero, in order to reduce the dimensionality of the parameter space that we need to explore. That being said, any large initial velocity, radial or angular, will be quickly red-shifted away \cite{Greenwood:2012aj}  making our choice natural.

We start by sampling  the initial conditions for the fields in two different ways:
 we consider the initial conditions to lie on a surface hypersphere  $\sum_I (\phi^I)^2 =R_{\rm sph}^2={\rm const}$ 
or a hyper-ellipsoid
$\sum_I \xi_I (\phi^I)^2 =R_{\rm ell}^2={\rm const}$. We found that, regardless of the choices of the potential parameters, sampling the initial field values on a hyper-ellipsoid leads to an extremely well-defined $e$-folding number. We know from the one- and two-field cases \cite{Kaiser:2013sna}, that when the system evolves along  one axis in field-space, the number of $e$-folds  is given by $N=\frac{3}{4}\xi \phi^2/M_{\rm {Pl}}^2$. We  generalize this to the many-field case, by considering the analytic formulas for the attractor 
given in Section~\ref{sec:attractor}.

\begin{figure}
    \centering    \includegraphics[width=0.32\textwidth]{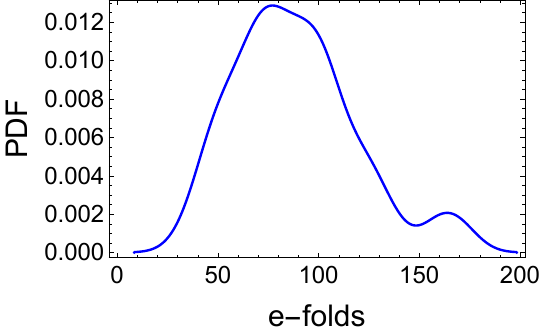 }
     \includegraphics[width=0.32\textwidth]{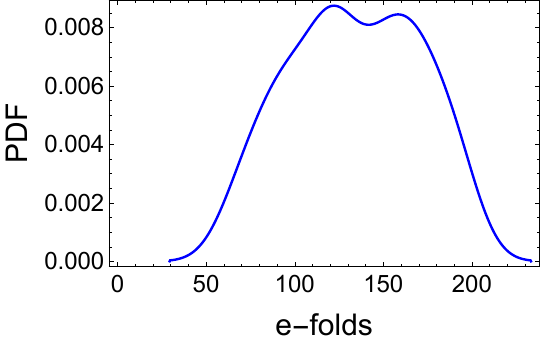 }
      \includegraphics[width=0.32\textwidth]{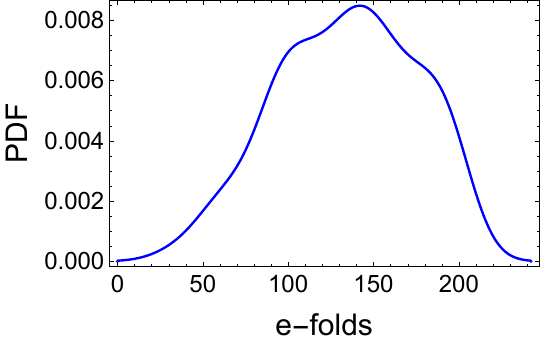 }
    \\
    \includegraphics[width=0.32\textwidth]{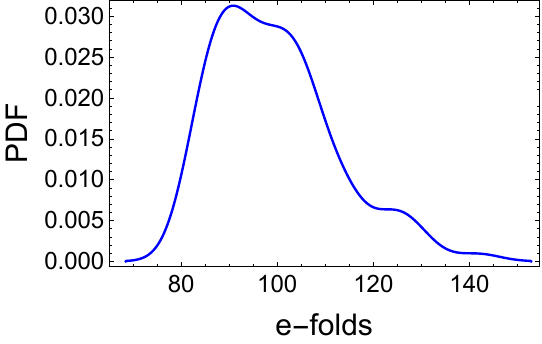}
        \includegraphics[width=0.32\textwidth]{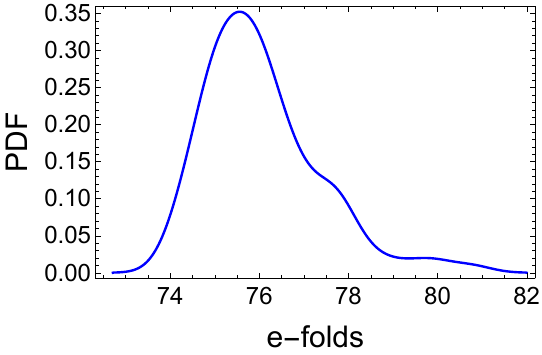}
    \includegraphics[width=0.32\textwidth]{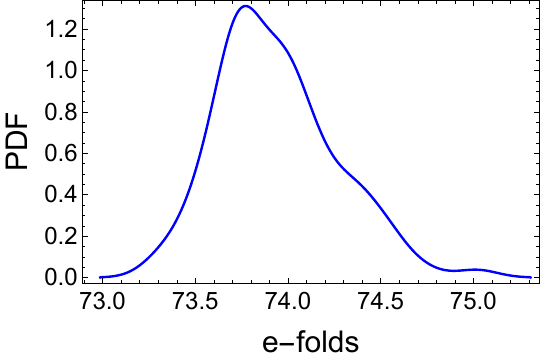}
    \caption{The number of $e$-folds of inflation for 1000 different runs, each chosen to lie on a hyper-sphere with constant $\sum_I  (\phi^I)^2$ (upper panels) or hyper-ellipsoid with constant $\sum_I  \xi_I (\phi^I)^2$ (lower panels) for $\xi={\cal O}(1, 10,100)$ from left to right. We see that the hyper-ellipsoid leads to very sharp distribution for the number of $e$-folds, which matches the analytical expectation for $\xi \ge {\cal O}(10) $. On the contrary, choosing fields on a hyper-sphere with constant $\sum(\phi^I)^2$ leads to a broad distribution of $e$-folds and thus very little predictive power.
    } 
    \label{fig:efoldnumber}
\end{figure}

In order to understand the dynamics involved, we performed simulations of systems with increased complexity, starting by choosing the initial conditions on a hypersphere or hyper-ellipsoid of different radii $R_{\rm sph}$ and $R_{\rm ell}$.
Unless otherwise stated, we set cross-couplings to zero, $g_{IJ}=0$. Figure~\ref{fig:efoldnumber} shows that choosing the initial conditions on a hyper-ellipsoid leads to a sharp distribution of $e$-folds for $\xi\gtrsim {\cal O}(10)$, in  agreement with the formula for the number of $e$-folds on the attractor
\beq
\label{eq:efoldnumber}
N  \simeq {3\over 4} \sum_I \frac{\xi_I (\phi^I)^2}{M_{\rm Pl}^2} \, .
\eeq
We can see both a shift and a spread from the above approximate analytical prediction. The shift is due to the transient phase that the system requires, until it reaches the single-field attractor. The spread has a different origin. The potential energy density of the system on the initial hyper-ellipsoid is not equal to the potential energy density of the system on the attractor for the same value of $R_{\rm ell}^2$. This leads to a spread on the number of $e$-folds, which is however much smaller than the average value of $N$, and also much smaller than the case of choosing initial conditions on a hyper-sphere, also show on Figure.~\ref{fig:efoldnumber}. In fact, all simulations with initial conditions lying on a hypersphere lead to a distribution of $e$-folding numbers with a standard deviation of ${\cal O}(10)$, whereas in the case of a hyper-ellipsoid the standard deviation is $\sim 1$ for $\xi=10$ and $\sim 0.25$ for $\xi=100$.


If we think of starting from ``generic'' initial conditions,  the hyper-ellipsoid that leads to the well-defined $e$-folding number wouldn't qualify. Instead, we need to choose  each field independently as a Gaussian random variable. This can be for example attributed to an earlier phase of (e.g.~eternal) inflation, when all fields accumulated de-Sitter fluctuations \cite{Starobinsky:1994bd}.

Armed with our understanding of the attractor behavior in these models, we can ask the question: How does the number of $e$-folds change when initial conditions and parameter values are drawn from different Probability Density Functions (PDFs).

\begin{figure}
    \centering    \includegraphics[width=0.45\textwidth]{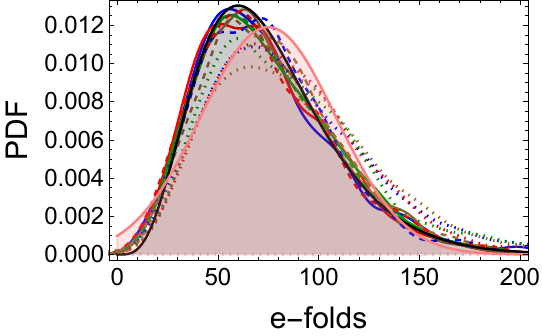}
    \centering    \includegraphics[width=0.45\textwidth]{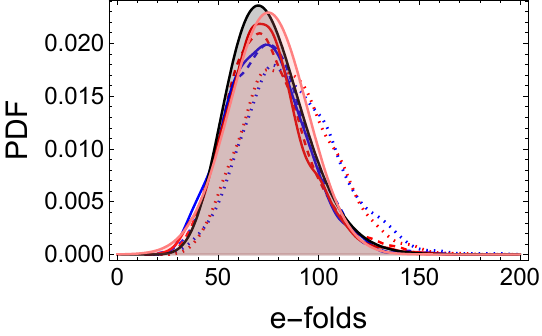}
    \caption{We show the resulting distributions arising from simulations with $10$ (left) and $30$  (right) fields.
    We show the $\chi^2$ and Gaussian distributions (grey-filled and red-filled respectively). The dotted, dashed and solid lines correspond to $\xi=1,10,100$ and the different colors correspond to different distributions of $\lambda$, uniform, log-uniform or Gaussian. We see that the distribution of $\lambda_I$ do not affect the $e$-folding number, as expected from our analysis. The amplitudes $\phi^I$ are drawn from a Gaussian distribution, whose variance is chosen, such that $\langle N\rangle\simeq 75$, following Eq.~\eqref{eq:chi2mean}.
    Finally, as expected, for larger number of fields, the $\chi^2$ and Normal distributions become increasingly more similar -- practically indistinguishable for ${\cal N}\gg 1$. We performed $1000$ simulations  for each curve.  }
    \label{fig:chi2}
\end{figure}

We model the initial condition for every field $\phi^I$  as an independent Gaussian (normal) random variable with zero mean and a given standard deviation.
It is worth noting here that as the number of fields increases, the details of the underlying distributions for the model parameters become less important, as the behavior of the system is mostly defined through their mean and variance. This is a consequence of the central limit theorem, as we will discuss later.
The final number of $e$-folds, according to our analytic approximation is given by Eq.~\eqref{eq:efoldnumber}.
If all non-minimal couplings are equal, the $e$-folding number can be re-written as
\beq
N = {3\over 4} \xi \sigma_\phi^2 \sum_{I=1}^{\cal N} {(\phi^I)^2\over \sigma_\phi^2 M_{\rm {Pl}}^2}
\eeq
where $\sigma_\phi^2$ is the variance of the distribution  of $\phi$, measured in units of $M_{\rm {Pl}}^2$. If we draw the field amplitude for each 
field $\phi^I$ from a normal PDF, the $e$-folding number is related to the  sum of
the squares of independent random variables with unit variance, which follows a $\chi^2$ distribution of order ${\cal N}$, thus
$N = {3\over 4} \xi^2 \sigma_\phi^2 y$, where $y\sim \chi^2_{\cal N}$:
\beq
\label{eq:chisquare}
f_{\chi^2}({y;{\cal N}}) = \frac{e^{-y/2} y^{\frac{\cal N}{2}-1} }{  
2^{{\cal N}/2} \Gamma({\cal N}/2)
}~,~ {\cal N}>0 \, .
\eeq
The mean and standard deviation are
\begin{eqnarray}
\label{eq:chi2mean}
\E(N) &=& {3\over 4} \xi\sigma_\phi^2 \E(y) = {3\over 4} \xi\sigma_\phi^2{\cal N}
\\
\label{eq:chi2var}
\sqrt{\Var(N)} &=&  {3\over 4} \xi\sigma_\phi^2   \sqrt{\Var (y)} = {3\over 4} \xi\sigma_\phi^2 \sqrt{2{\cal N}}
\end{eqnarray}
We must stress here that choosing all non-minimal couplings equal does  not lead to a radially symmetric potential or even to a potential that is identical in all directions $\phi^I$. We still randomly choose all self-couplings $\lambda_I$, thus the direction of the attractor is randomly distributed along the angles of the ${\cal N}$-sphere with the relations $\phi^I/\phi^J = \sqrt{\lambda_J / \lambda_I}$.
Figure~\ref{fig:chi2} shows the comparison of the numerical results with the analytical approximation of Eq.~\eqref{eq:chisquare}. In the case of many fields, we can use the central limit theorem, which means that the $\chi^2$ distribution approaches a Gaussian with mean ${\cal N}$ and variance $2{\cal N}$. 
Interestingly, the mean and variance of the $\chi^2$ distribution are equal to the Gaussian approximation. However, as seen in Figure~\ref{fig:chi2}, the peak of the $\chi^2$ is always to the left of the mean of the corresponding Gaussian distribution, specifically the peak of $\chi^2_{\cal N}$ occurs at ${\cal N}-2$. For ${\cal N}\gg 1$ the asymmetry of the distribution is hardly visible.
For $\sigma_\phi=1$, we arrive at $\mu_N=75, 750$ for $\xi=10,100$ and $\sigma_N = 33.5, 335$ for $\xi=10,100$. 
The left panel of Figure~\ref{fig:chi2} shows the $\chi^2$ and normal distributions, together with the PDFs for $10$ fields and $10^3$ runs. For $\xi=10,100$, it is clear that our simulations follow the distribution $\chi^2$ very closely and are visibly different from the corresponding Gaussian, derived through the central limit theorem. For $\xi=1$ the match is not as clear, since $\xi=1$ does not fall within the large $\xi$ regime, where the attractor equations can be applied. Nevertheless, even for $\xi=1$, the mean and variance of the PDF for the number of $e$-folds still follows our predictions well enough to provide simple estimates. 
For more fields, the difference between the Gaussian and $\chi^2$ distributions is practically negligible and certainly smaller than the inherent variance of our simulation, as shown in the right panel of Figure~\ref{fig:chi2}.

The most important outcome of this numerical experiment is the complete insensitivity of the distribution of $e$-folds on the choice of $\lambda_I$.  Regardless of the form of the PDF that defines the distribution of the self-couplings, the resulting distribution of $e$-folding numbers is the same. This further reinforces our analytical understanding of the single-field attractor.

\begin{figure}
    \centering
\includegraphics[width=.45\textwidth]{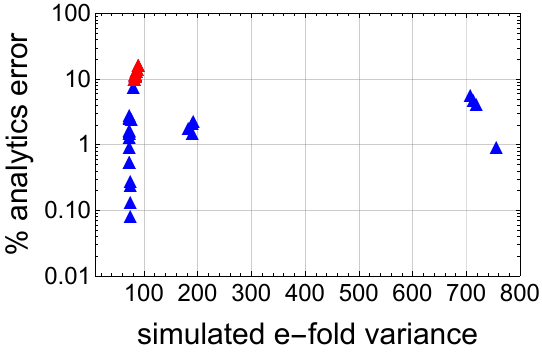}
\includegraphics[width=.45\textwidth]{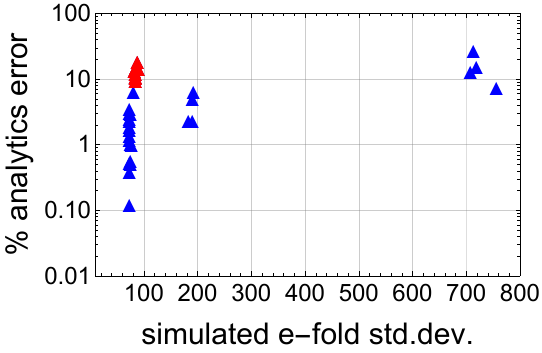}
    \caption{ The mean (left) and standard deviation (right) of the number of $e$-folds of inflation, computed using the simulated data and the relative percentage difference between simulated data and the mean / standard deviation computed using Eqs.~\eqref{eq:average_efolds_analytics}, \eqref{eq:stdev_efolds_analytics}. We see the agreement to be better than $10\%$ for most cases. The points in red correspond to $\xi={\cal O}(1)$ and exhibit the larger percentage difference in the mean number of $e$-folds as expected, since the single field attractor is weaker for $\xi\lesssim 1$.
 }
    \label{fig:efoldscatter}
\end{figure}

If the non-minimal couplings are also drawn from a distribution,  we need to first define the PDF of the random variable $\xi_I(\phi^I)^2$ as a product distribution, before evaluating the sum of Eq.~\eqref{eq:efoldnumber}. We keep drawing $\phi^I$ from a normal distribution, but we allow ourselves more freedom to draw the values of $\xi_I$. Because of this, the product distribution will not have a simple analytical expression in general.\footnote{A product distribution $Z=XY$ follows a PDF with $f_Z(z) = \int_{-\infty}^\infty dx\, \frac{1}{|x|}f_X(x)f_Y(z/x)$.} However, we can still compute the properties of it, based on the formulae for the mean and variance of a product of independent random variables. 
\begin{eqnarray}
\label{eq:average_efolds_analytics}
\E \left [ \xi \phi^2 \right ] &= & \E\left [\xi\right ] \E\left [\phi^2\right ] \equiv \mu_\xi \mu_{\phi^2} 
=\mu_\xi \sigma_\phi^2
\\
\label{eq:stdev_efolds_analytics}
{\Var}\left [\xi\phi^2\right ] &= & \left (\sigma_\xi^2 + \mu_{\xi}^2\right)
\left (\sigma_{\phi^2}^2 + \mu_{\phi^2}^2 \right ) -\mu_{\xi}^2 \mu_{\phi^2}^2 \, ,
\end{eqnarray}
where the mean $\E(\xi)\equiv \mu_\xi$ and variance ${\rm {Var}}[\xi] \equiv \sigma_\xi^2$ of the $\xi$ distribution  depend on the specific choice of the PDF and the corresponding  mean and variance of the distribution of $\phi^2$ are computed using the $\chi$-squared distribution, as discussed above. 
Fortunately, the use of the central limit theorem does not require full knowledge of the PDF of each random variable. This means that, in the limit of large number of fields ${\cal N}\gg 1$, we can approximate the result of $\sum_{i=1}^{\cal N} \xi_I (\phi^I)^2$ as a Gaussian with mean ${\cal N} \E \left [ \xi \phi^2 \right ]$ and variance ${\cal N}{\rm {Var}}\left [\xi\phi^2\right ] $. 
Figure~\ref{fig:efoldscatter} 
 shows the mean and standard deviation of the resulting $e$-folding number distribution for difference choices for the distributions of the parameters and initial conditions. We see that the analytical and numerical values of $\mu_{\cal N}$ and $\sigma_{\cal N}$ are in excellent agreement with each other for a wide range of simulations and parameter choices. More importantly, the distribution of the self-couplings does not affect the distribution of $e$-folding numbers, as expected from our analytic results. However, the Hubble scale does depend on $\lambda_I$, as we demonstrate below.

\bigskip

\subsection{Hubble scale and turn-rate}

We can further utilize the strength of the single field attractor and the effectiveness of our analytic results, by examining the distribution of the Hubble parameter and turn rate for different realizations of the model. Assuming again that the cross-couplings vanish, $g_{IJ}=0$, we recover the simple form of the Hubble scale during inflation
\beq
H^2 = \frac{M_{\rm {Pl}}^2}{12} \frac{1}{\sum_{I=1}^{\cal N}\left ({\xi^2_I} /{ \lambda_I} \right )}
\label{eq:hubbleattractor}
\eeq

There are several steps in computing the mean and variance of the distribution of Hubble parameters for different realizations. 
 Given independent PDFs for $\xi_I$ and $\lambda_I$, both of which take only positive values, the expectation value of this ratio random variable can be computed using the moments of the original distributions
$\E\left [\xi^2/\lambda\right ] = \E[\xi^2]\E[\lambda^{-1}]$ and similarly $E\left [(\xi^2/\lambda)^2\right ]= \E[\xi^4]\E[\lambda^{-2}]$, which are straightforward to compute given concrete forms of the PDFs for $\xi,\lambda$.
 The sum (defined  
 as $y = \sum_I \xi_I^2/\lambda_I$), in the limit of many fields, approaches a normal distribution, with mean ${\cal N}\tilde \mu$ and standard deviation $\sqrt{\cal N}\tilde \sigma$, where $\tilde\mu,\tilde\sigma$ are the corresponding quantities for the random variable ratio $\xi^2/\lambda$.
Finally, the distribution of Hubble parameters approaches a Reciprocal Normal distribution (in the regime of validity of the central limit theorem). Specifically, with $z=1/y$, the random variable $z$ follows the PDF 
\beq
f(z) = \frac{1}{\sqrt{2\pi}\sigma z^2} e^{-{1\over 2} \left ({1/z -\mu \over \sigma}\right )^2}\simeq
{\mu_y^2  \over \sqrt{2\pi} \sigma_y} e^{-{\mu_y^4\over 2\sigma_y^2}\left ( {z-{1\over \mu_y}} \right )^2} \, ,
\eeq 
where $\mu_z \simeq 1/\mu_y$ and $\sigma_z \simeq \sigma_y^2/\mu_y^4$. The derivation for the above formula is given in Appendix~\ref{app:inversedist}.

\begin{figure}
    \centering
\includegraphics[width=.45\textwidth]{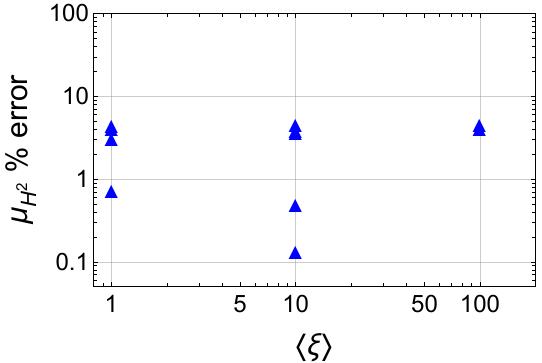}
\includegraphics[width=.45\textwidth]{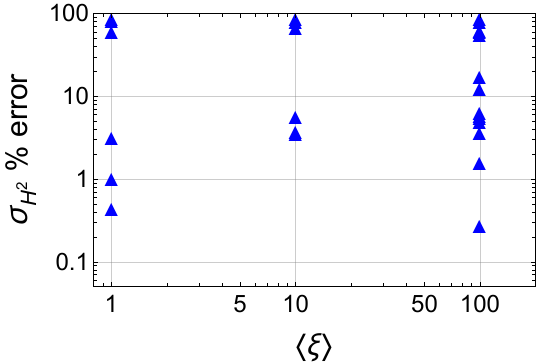}
    \caption{ The percentage error of the mean (left) and standard deviation (right) for  $H^2$ (proportional to the energy density during inflation) at $N=50$ $e$-folds before the end of inflation, computed using the simulated data and the mean and variance of the input distributions. 
    We see that the mean is computed very accurately using our approximation, whereas the standard deviation can have (at most) a factor-$2$ difference for a few cases, but is still at the $10\%$ level or better for most cases.
    }
    \label{fig:Hubblescatter}
\end{figure}

We have thus set up a procedure to compute the PDF of the Hubble scale, given PDFs for $\xi$ and $\lambda$. The result is independent of the initial field values $\phi^I$, as expected from the evolution of the system along the attractor. Figure~\ref{fig:Hubblescatter} shows the estimated and simulated mean and standard deviation for a wide range of simulations, again showing very good agreement, within the range of validity of our approximation. It is worth noting that even in the single-field case, the analytic approximation differs from the exact result by ${\cal O}(1\%)$ for $\xi\gg 1$ and more for $\xi={\cal O}(1)$.

\begin{figure}
\includegraphics[width=.45\textwidth]{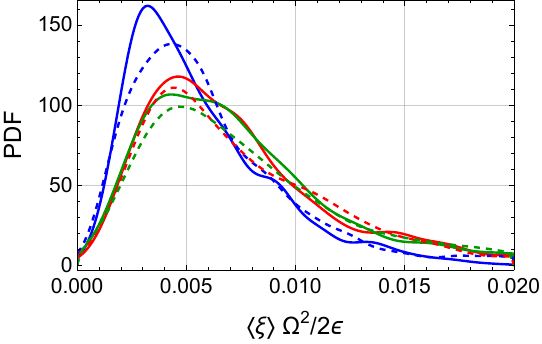}
\includegraphics[width=.45\textwidth]{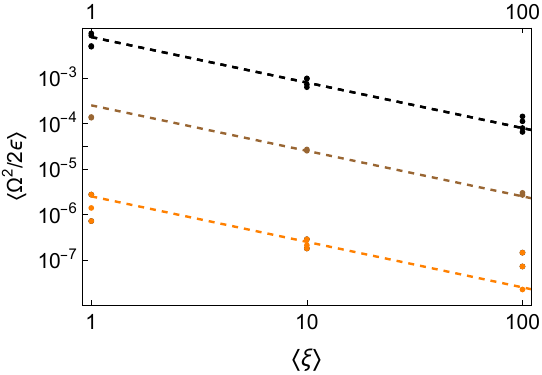}
\caption{
{\it Left:} The rescaled normalized turn-rate for differnt runs with $\xi$ distributions centered around $1$ (blue), $10$ (red) and $100$ (green). The solid and dashed curves correspond to different choices of initial conditions (lying on a hyper-sphere or hyper-ellipsoid respectively).
{\it Right:} The mean of the distribution of the normalized turn rate $\Omega^2/2\epsilon$ as a function of the non-minimal coupling $\langle \xi\rangle$ for different runs. The scaling $ \Omega^2/2\epsilon \propto 1/ \xi $ is given by the  dashed lines. Different colors correspond to different batches of runs, with similar choices for the parameter distributions. We see that the overall magnitude (pre-factor) of the turn rate depends strongly on the chosen distributions, whereas the scaling with $\xi$ is clear and universal for each batch of runs. }
\label{fig:omega_distribution}
\end{figure}

In Section~\ref{sec:attractor} we
provided novel results for the normalized turn rate $\Omega$ of a multi-field system with non-minimal couplings during inflation, culminating in the relation $\Omega^2 \propto {\epsilon /\xi}$, where the proportionality factor is calculable (albeit complicated) and contains ratios of the form $\xi_I/\xi_J$, $\lambda_I/\lambda_J$ and is found to be (much) smaller than unity in every case we examined. In Figure~\ref{fig:omega_distribution} we plot the quantity $\Omega^2 / ({2\epsilon})$ for
the simulations contained in Figure~\ref{fig:efoldnumber}, where we choose the initial conditions for the fields lying on a hyper-sphere and hyper-ellipsoid and choose $\lambda$ and $\xi$ from distributions with $\langle \xi\rangle =1,10,100$. Intriguingly, the distributions of the rescaled normalized turn-rate for the simulations with $\langle \xi\rangle \ge 10$ are essentially identical. The two simulations for $\langle \xi\rangle =1 $ are almost equal to each other but slightly different from the ones for larger values of $\xi$. This is expected since Eqs.~\eqref{eq:asymptotic} and \eqref{eq:asymptotic2} are valid for large values of $\xi$. However, the right panel of Figure~\ref{fig:omega_distribution} shows that the mean normalized turn rate follows the scaling $\Omega^2/2\epsilon \propto 1/\langle \xi\rangle$, albeit with a prefactor that can change significantly for different parameter distributions.

\subsection{General case}
\label{sec:Generalcase}

\begin{figure}
    \centering  
    \includegraphics[width=0.4\textwidth]{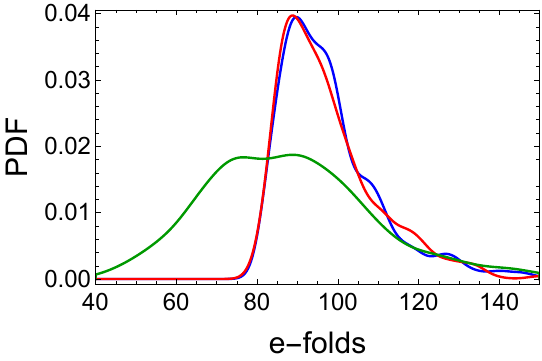}\quad\quad
        \includegraphics[width=0.4\textwidth]{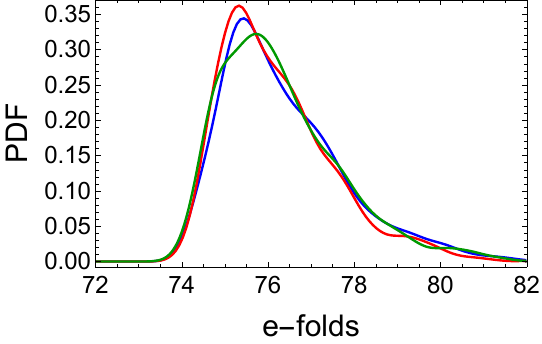}
        \\
         \includegraphics[width=0.4\textwidth]{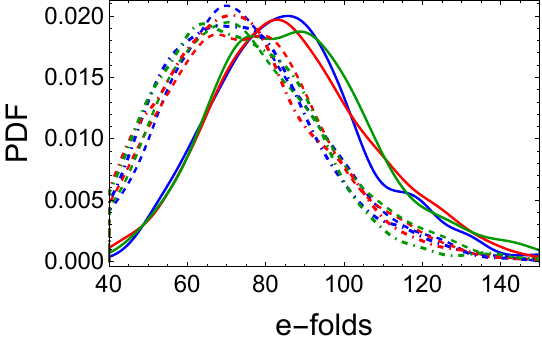}\quad\quad
    \includegraphics[width=0.4\textwidth]{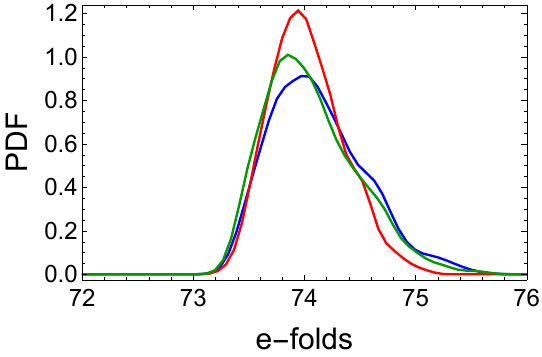}
\caption{
The number of $e$-folds of inflation for  different runs, each chosen to lie on a hyper-ellipsoid with constant $\sum_I  \xi_I (\phi^I)^2$ for $\xi={\cal O}(1, 10,100)$ clockwise from top left. 
The red curves correspond to $g_{IJ}>0$ with different distributions, the blue curves correspond to $g_{IJ}$ taking positive and negative values and the green curves correspond to $g_{IJ}=0$. 
We see that choosing initial conditions on a hyper-ellipsoid leads to very sharp distribution for the number of $e$-folds, with $g_{IJ}$ affecting the distribution of $e$-folds slightly.
The bottom left panel corresponds to the same parameter choices but with fields initialized on a hypersphere of constant $\sum_I(\phi^I)^2$ for $\xi\sim 1,10,100$ (solid, dashed and dot-dashed respectively).
}
\label{fig:efold_gIJ}
\end{figure}

The analysis so far has been limited to the case of zero $g_{IJ}$, in order to reduce the number of free parameters. However, the results do not change significantly, even if we allow for random $g_{IJ}$. 
As shown in Appendix~\ref{app:attractor}, the inclusion of attractive cross-couplings $g_{IJ}<0$ can result in the potential turning negative.
In order to avoid this, 
we can choose cross-couplings from a distribution with strictly non-negative values. Fig.~\ref{fig:efold_gIJ} shows the resulting distribution for the number of $e$-folds for $10$ fields and $1000$ different simulations, when choosing the initial field values to lie on a hyper-ellipsoid $\sum_I \xi_I (\phi^I)^2 = {\rm {const}}$ or a hyper-sphere $\sum_I (\phi^I)^2 = {\rm {const}}$. We see that the resulting distribution for the hyper-ellipsoid is still sharply peaked around the theoretical value for $\xi\gg 1$. Since restricting ourselves solely to repulsive couplings $g_{IJ}>0$ can be considered fine-tuned, we performed a similar set of $1000$ runs, where we multiplied the $g_{IJ}$ with $\pm 1$ with $50\%$ probability. 
Interestingly, when we choose $\lambda_I$ and $|g_{IJ}|$ from the same Marchenko-Pastur distributions, the potential exhibits negative values (leading to the runs terminating in $V<0$ instead of $\epsilon=1$) about $90\%$ of the time. By reducing the magnitude of $g_{IJ}$ by one order of magnitude compared to $\lambda_I$, the ``failed" runs were limited to about $20\%$.\footnote{When choosing $\lambda_I$ and $|g_{IJ}|$ from a uniform distribution the failure rate drops to less than $5\%$ when we reduce the magnitude of  $g_{IJ}$ by one order of magnitude compared to $\lambda_I$. This is due to the shape of the  Marchenko-Pastur distribution (with its wide ``tail") and the fact that only one very negative large cross-coupling  is needed for the potential to turn negative.} We continue drawing new samples, until  we reach $1000$ successful runs to achieve the same level of statistics. The resulting distribution of $e$-folding numbers is shown in Fig.~\ref{fig:efold_gIJ}. The peaked distribution is still evident, regardless of $g_{IJ}$. In fact, for $\xi\gg 1$, the $e$-folding distribution has a standard deviation of less than one $e$-fold, regardless of the choice made for $g_{IJ}$.
Overall, we see that the single-field-like attractor of these models is extremely robust. Since all other results of our work (e.g. the distribution of Hubble values) are derived using the attractor, these results can be mapped onto  the general case $g_{IJ}\ne 0$. Even though the analytical results will be more difficult to obtain (requiring   the eigenvalues of large matrices), the physical intuition and scaling arguments carry over.

In order to describe the background dynamics as a function of different parameter choices in a more intuitive way, we present the distribution of background quantities ($e$-folding number, Hubble scale and turn rate) for $11$ different cases, each one differing from the ``benchmark" case by only one parameter choice. The benchmark case A consists of a system of 
$10$ fields with zero cross-couplings $g_{IJ}=0$, non-minimal couplings drawn from a Gaussian distribution with mean $\langle \xi\rangle = 50$ and standard deviation $\sigma_\xi=10$, quartic self-couplings drawn from a uniform distribution $\lambda \in [.01,1]$ and initial field values drawn from a normal distribution with zero mean and variance 
$\sigma_\phi= 0.5 M_{\rm {Pl}}$, chosen to produce around $100$ $e$-folds of inflation. 
For each further pair of runs, we change only one of the parameter distributions. In particular,
\begin{itemize}
\item For cases B and C we change the standard deviation of the initial conditions PDF by factors of $2$ and $0.5$ respectively.

\item 
For cases D and E, we change the upper limit of the $\lambda $ PDF by factors of $10$ and $0.1$, respectively. 

\item For cases F and G, we change the standard deviation of the $\xi$ PDF by factors of $2$ and $0.5$, respectively, while keeping the mean fixed.

\item For cases H and I, we change the mean of the $\xi$ PDF by factors of $2$ and $0.5$, respectively, while keeping the standard deviation fixed.

\item Finally, for case J and K we choose cross-couplings $g_{IJ}$ drawn from a uniform distribution with or without a random sign, respectively.
\end{itemize}

\begin{figure}
    \centering  
    \includegraphics[width=0.36\textwidth]{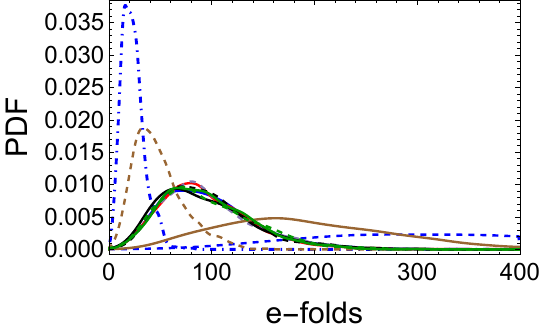}
        \includegraphics[width=0.5\textwidth]{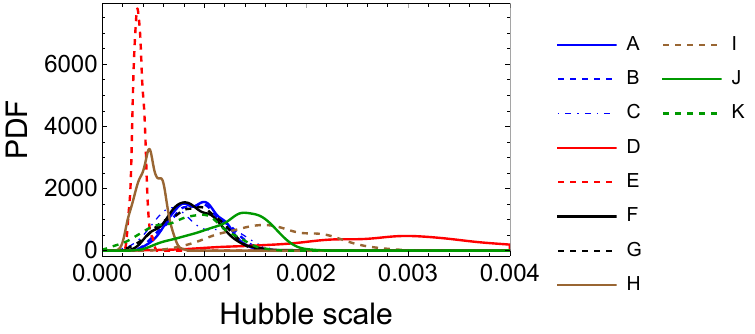}
        \\
     \includegraphics[width=0.36\textwidth]{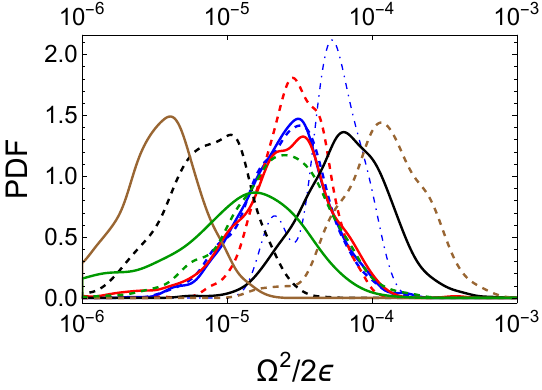}
        \includegraphics[width=0.5\textwidth]{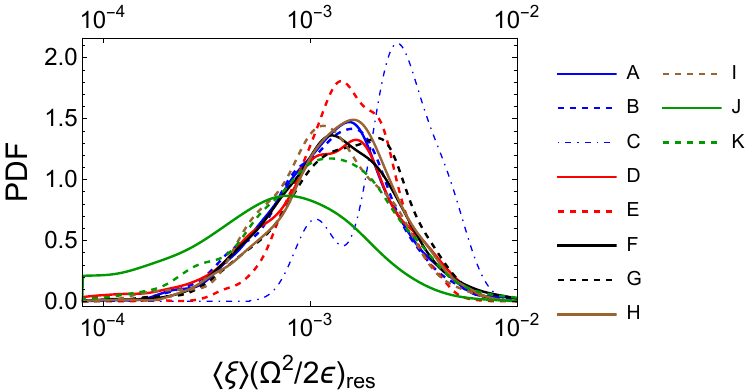}
\caption{
{\it Upper panels:} The distribution of the $e$-folding number (left) and Hubble scale (right).
{\it Lower panels:} The distribution of the normalized turn rate $\Omega^2/2\epsilon$ (left) and  normalized turn-rate $\langle \xi\rangle \Omega^2/2\epsilon$ rescaled by $R_\xi$ as discussed in the main text.
}
\label{fig:summary_cases}
\end{figure}

Figure~\ref{fig:summary_cases} shows the PDFs for the $e$-folding number, the Hubble scale, and the turn rate normalized in two different ways ($\Omega^2/2\epsilon$ and $\langle \xi\rangle (\Omega^2/2\epsilon)_{\rm {res}}$)
for the different cases A-K.
We first number of $e$-folds of inflation, where the resulting distribution from almost all cases is identical. An exception is given by cases B and C, where we changed the variance of the initial field amplitude, which changes $N$ correspondingly, according to our analysis of the single field attractor, following Eqs.~\eqref{eq:average_efolds_analytics} and \eqref{eq:stdev_efolds_analytics}. Furthermore, Eq.~\eqref{eq:average_efolds_analytics} shows that the mean $e$-folding number is proportional to the mean value of the non-minimal couplings $\langle \xi\rangle$. This is also seen in the brown curves in Fig.~\ref{fig:summary_cases}. The variance of $N$ is proportional to $\mu_\xi^2 + \sigma_\xi^2$. However, in the examples we chose $\sigma_\xi>\mu_\xi$ and in particular the contribution of $\sigma_\xi^2$ to the above sum is at most $16\%$. Thus, changing the variance of the $\xi$ distribution does not alter the results, as long as the condition $\sigma_\xi^2\ll \mu_\xi^2$ is satisfied. 

The top right panel of Fig.~\ref{fig:summary_cases} shows the Hubble scale for cases A-K. Examining the formula for the Hubble scale along the attractor, Eq.~\eqref{eq:hubbleattractor}, allows for an intuitive understanding of the different curves.
Most curves are identical to case A, since for example the initial values of the fields do not affect the Hubble scale. There are three cases in which a large deviation is seen. We first look at the red curves, corresponding to cases D and E, where the distribution of the quartic self-couplings $\lambda$ is changed. Since  $\lambda$ 
 sets the scale of the potential, increasing the overall value of $\lambda$ increases the value of $H$ and vice versa. This is shown in Figure~\ref{fig:summary_cases}. Similarly, changing the scale of the non-minimal couplings has a similar (but opposite) effect on $H$, which is shown in the brown figures (cases H and I), as expected. A separate comment must be made about cases J and K, where we introduce cross-couplings between the fields. If the cross-couplings are all positive (case K), the Hubble scale is largely unaffected, whereas if we allow negative values, we see a (slightly) larger Hubble scale. 

 Finally, we examine the two lower panels of Figure~\ref{fig:summary_cases}, which show the distribution of the turn rate, normalized in different ways. The left panel shows the simple expression $\Omega^2/2\epsilon$ with no further rescaling. We see that only the blue and red curves overlap. The blue curves correspond to changing initial values of the fields, which does not affect the turn rate, as expected\footnote{We only include the turn-rate for runs that exceed $60$ $e$-folds of inflation in the distributions that we show in Figure~\ref{fig:summary_cases}. For run C, this leaves less than $10$ runs and thus very low statistics, not allowing us to draw conclusions for the turn rate based on the blue dot-dashed curve.} The red curves, corresponding to changing the width of the uniform distribution of $\lambda$ are also very similar to the blue ones, due to the simple behavior of $R_\lambda$, approaching a uniform distribution in the range $\{0,1\}$,  when $\lambda$ is drawn itself from a uniform distribution. All other curves differ significantly\footnote{Of the green curves, one overlaps with case A, whereas the other doesn't. We expect cases with $g_{IJ}\ne 0$ to introduce additional terms similar to $R_\lambda$ in Eqs.~\eqref{eq:asymptotic}, \eqref{eq:asymptotic} and their many-field counterparts. However, we do not pursue a detailed analytic treatment of this case here.}. In order to check whether the ``universal" distribution of the rescaled turn rate holds in the many-field case, we first multiply by $\langle \xi\rangle$. However, this does not produce the expected universality. However, examining Eqs.~\eqref{eq:asymptotic} and \eqref{eq:asymptotic2}, we see that in both the two and three field cases, the turn rate scales as $\Omega^2/2\epsilon \propto (1-R_\xi)^2$ where $R_\xi<1$ is the ratio of two non-minimal couplings, drawn from the relevant distribution. By computing this for the uniform distribution at hand, we arrive 
 at $\bar R_\xi \equiv \langle R_\xi \rangle \simeq0.04$ for case A, $\bar R_\xi\simeq0.1$  cases F and I and $\bar R_\xi\simeq0.01$ for cases G and H. The right panel of Figure~\ref{fig:summary_cases} shows the turn rate, multiplied by $\langle \xi\rangle$ and with the extra rescaling $(\Omega^2/2\epsilon)_{\rm{res}} \equiv 
\left [(1 - \bar R_\xi^A)^2 /  (1 - \bar R_\xi^i)^2\right ](\Omega^2/2\epsilon)$ for each case $i=$A,B,C etc. We see a remarkable universality, where all distributions fall largely on top of each other, with the exception of case C, whose low statistics make it irrelevant, and case $J$, where the existence of non-zero cross-couplings $g_{IJ}$ would introduce extra terms similar to $R_\lambda$, thereby shifting the distribution by an ${\cal O}(1)$ number. This further showcases the strength of the inflationary attractor and the ability of our analytical treatment to capture its statistical properties for all relevant background quantities.

\subsection{CMB Observables}

We have demonstrated the strength of the single field attractor, as well as its surprising simplicity, which allows us to describe all background dynamical quantities, the number of $e$-folds, the Hubble scale and the turn rate as simple distributions with calculable mean and standard deviation. 
However, in order to fully understand the observables arising from a family of inflationary models, we also need to study the fluctuations.

\begin{figure}
    \centering  
    \includegraphics[width=0.45\textwidth]{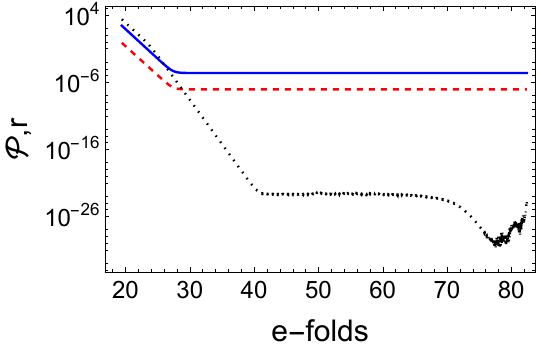}
        \includegraphics[width=0.48\textwidth]{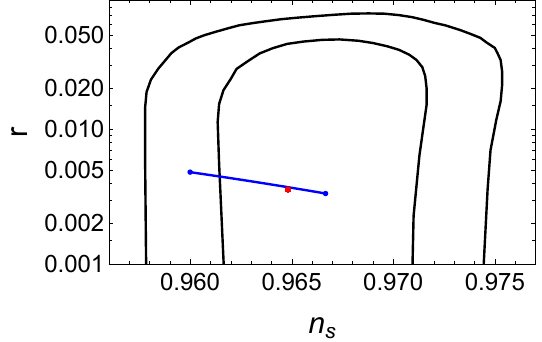}
      
\caption{
{\it Left:} The adiabatic (blue) and isocurvature (black-dotted) power for a mode exiting the horizon $55$ $e$-folds before the end of inflation  for a single realization of the system with $10$ fields. The red-dashed curve shows the power in tensor modes. 
{\it Right:} The predictions for $n_s$ and $r$ (red dots) along with the $1\sigma$ and $2\sigma$ curves, showing joint constraints from  {\it Planck} with BAO and Bicep/Keck~\cite{Planck:2018vyg}. There are $45$ different dots on the plot, even though the attractor behavior makes them largely indistinguishable. The blue line corresponds to the analytic slow roll predictions for $n_s$ and $r$ for the CMB modes exiting the horizon between $60$ and $50$ $e$-folds between the end of inflation. The numerical results were computed using $55$ $e$-folds.
}
\label{fig:nsr}
\end{figure}

Given the smallness of the turn rate during the slow-turn phase, we expect the observables to be indistinguishable from single-field ones. We only expect deviations when part of the last $50-60$ $e$-folds lies on the transient phase, or when the system rolls on the ridge for an extensive period when isocurvature perturbations have enough time to amplify. For the number of fields (up to $10$) and the parameter distributions we used (with large enough spread in the self-couplings and / or non-minimal couplings), such trajectories did not materialize with a rate that would show up in our analysis.\footnote{We should note that for all runs, we required around $100$ $e$-folds of inflation, so that the system has settled to the attractor by $N_*=60$. We consider the opposite to be a fine-tuned scenario, which in our statistical framework carries very low probability and is thus neglected.} For single-field non-minimally coupled inflation, the well-known relation arises $n_s\simeq 1-2/N_*$ where $N_*$ is the number of $e$-folds before the end of inflation where the modes corresponding to the CMB pivot scale exited the horizon. 
This translates to $n_s= 0.966$ and for $N_* = 60$ and $n_s = 0.959$ for $N_* = 50$, which was further verified for two (and three) field models \cite{Kaiser:2013sna}, with the numerical results agreeing with the simple slow-roll formula to the percent level or better. Our numerical results for the $10$-field runs similarly agree with the expectation from the existence of a single-field attractor. We also find the tensor-to-scalar ratio to be $r\simeq 3\times 10^{-3}$ and the running of the spectral tilt to be $\alpha = -{\cal O}(10^{-4})$, as expected \cite{Kaiser:2013sna}. Finally, we observed that neither isocurvature modes nor non-Gaussianity are produced at CMB scales. The results are shown in Figure~\ref{fig:nsr}



\section{Summary and Conclusions}
\label{sec:summary}

In the current work, we revisited the well-motivated models of inflation driven by scalar fields coupled non-minimally to gravity. Going beyond 
previous analyses,
we explored the many-field regime 
using both analytical and numerical techniques. 
We discovered the existence of a single field attractor, which is described by very simple formulae for its direction in field-space as well as the Hubble scale. 

The existence of a strong attractor allowed us to derive the statistical properties of all background quantities, including the number of $e$-fields, the Hubble scale, and the turn rate, as a function of the statistical properties of the couplings and initial conditions. We performed simulations with samples drawn from different distributions, and the results matched our analytical expectations.
Interestingly, the turn rate for an attractor that does not evolve along only one of the field directions is not zero, and was captured by our analysis. However, the resulting CMB observables, in particular the adiabatic spectral tilt $n_s$ match the single-field values, due to the smallness of the turn rate along the attractor and the large value of the corresponding isocurvature mass.
In the case of a softly broken symmetry, where no hierarchy exists between the different non-minimal couplings, the attractor formally exists, but can be evaded by the evolution of the system for a prolonged period. This will be presented separately in a subsequent publication.

\acknowledgments{}
We thank David Kaiser for useful discussions.
RR and EIS gratefully acknowledge support from the Simons Center for Geometry and Physics, Stony Brook University at which some 
of the research for this paper was performed.
PC is supported in part by the
National Research Foundation of Korea Grant 2019R1A2C2085023.  
EIS acknowledges support of a fellowship from ``la Caixa'' Foundation (ID 100010434) and from the European Union's Horizon 2020 research and innovation programme under the Marie Sklodowska-Curie grant agreement No 847648. The fellowship code is LCF/BQ/PI20/11760021. PC thanks the Korea Institute for Advanced Study for its hospitality.
RR completed much of this work while appointed to the NASA Postdoctoral Program at the NASA Marshall Space Flight Center, administered by Oak Ridge Associated Universities under
contract with NASA.

\appendix

\section{Full analytics on attractor behavior}
\label{app:attractor}


\subsection{Vanishing cross-couplings} \label{sec_vanishing_cc}
To study the dynamics of the problem we first ignore the geometry and focus on the potential, assuming gradient-flow evolution. 
Performing the coordinate transformation 
\begin{align}
\phi^K = r \sqrt{\xi_{(K)}} x^{(K)} \, , \qquad \sum_{I=1}^{\mathcal{N}} (\phi^I)^2 = r^2 \, , \qquad \sum_{I=1}^{\mathcal{N}} (x^I)^2 = 1 \, ,
\end{align}
(no sum in $K$) the potential becomes
\beq \label{eq:potential_r_xi}
V(r,x^I) =  {    r^4 \over (1 + r^2)^2}  \sum_{I=1}^{\mathcal{N}}   {\lambda_I \over \xi_I^2 } x_I^4 =  {    r^4 \over (1 + r^2)^2}  \sum_{I=I}^{\mathcal{N}}   {1 \over \zeta_I^2 } x_I^4  \, ,
\eeq
where in the last expression we absorbed the self-couplings $\lambda_I$ with a redefinition of the non-minimal couplings given by 
$
\zeta_I \equiv \xi_I/\sqrt{\lambda_I} \, .
$
The problem is therefore equivalent to a product-separable potential $V=F(r)W(x^I)$ where the $W$ part is sum of constrained quartic potentials with self-couplings $\zeta_I^{-2}$. For this model, $r$ cannot be the heavy field because its potential has critical points only at $r=0$ which is the global minimum of the potential. We conclude that $r$ is the light field and $x^I$ are the heavy fields.

\begin{figure}
\includegraphics[width=.45\textwidth]{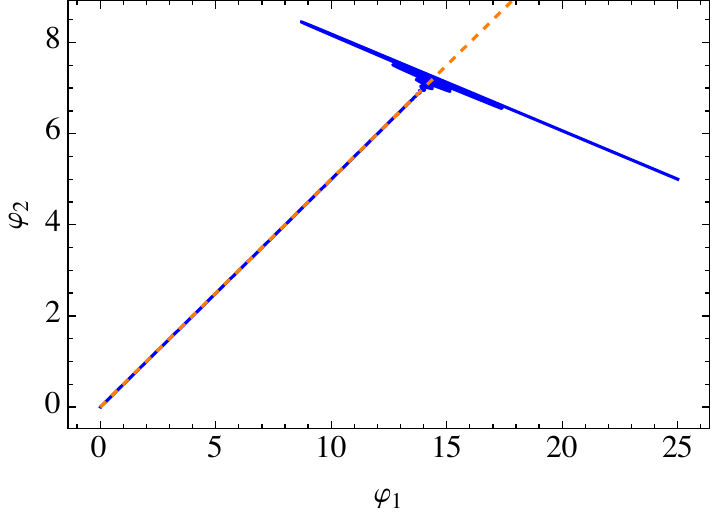}
\includegraphics[width=.45\textwidth]{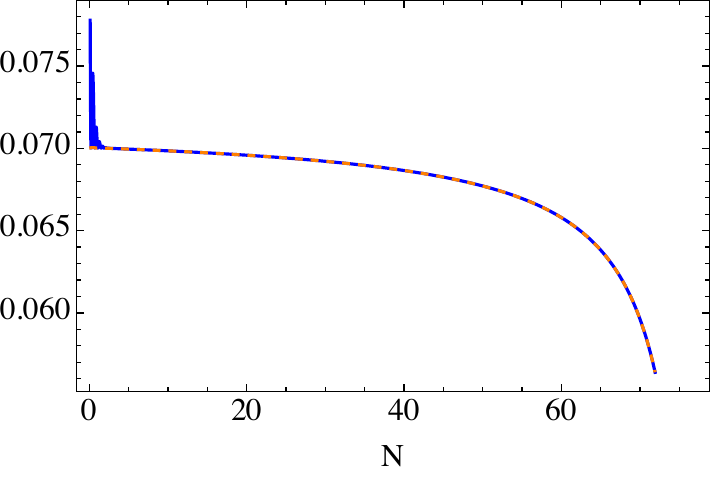}
\caption{Numerical (blue) versus the attractor expressions (orange, dashed) for a two-field realization with $\lambda_1=1,~\lambda_2=10,~\xi_1=2,~\xi_2=10$: (Left) The parametric relation between the fields compared to \eqref{eq:attractor_expression_ij}; (Right) The potential versus its attractor expression \eqref{eq:potential_attractor}.}
\label{fig:van_cross_coupl_equal_couplings}
\end{figure}

The attractor solution will be given at the values of $x^I$ for which $V_{,x^I}=0$. Note that only $\mathcal{N}-1$ fields are independent and we can parametrize e.g.~the last of them as
\beq
 x^{\mathcal{N}} = \pm \sqrt{1- \sum^{\mathcal{N}-1}_{I=1} (x^I)^2} \, .
\eeq
For the coordinate transformation to be consistent we exclude the case $x^{\mathcal{N}}=0$. The critical points of the potential for the heavy fields satisfy
\beq \label{eq:critical_point_effective_gradient}
{(x^K)^2 \over \zeta_K^2 }= {\left(x^{\mathcal{N}}\right)^2 \over \zeta_{\mathcal{N}}^2}  \, ,
\eeq
which holds for arbitrary $K$ and hence the ratio $x^K / \zeta_K$ is constant for all $K$. The previous relation implies that 
\beq \label{eq:sum_zetak}
\sum_{I=1}^{\mathcal{N}}  \zeta_{K}^2 = { \zeta_{\mathcal{N}}^2  \over \left(x^{\mathcal{N}} \right)^2 } 
\eeq
and thus the heavy fields are stabilized at values equal to
\beq  \label{eq:zk}
\left (x^K\right )^2 = {\zeta_{K}^2 \over \sum_{I=1}^{\mathcal{N}}  \zeta_{I}^2} \, .
\eeq
Multiplying Eq.~\eqref{eq:critical_point_effective_gradient} with $\left(x^I\right )^2$ and summing yields
\begin{align}
\sum_{I=1}^{\mathcal{N}}  {(x^I)^4 \over \zeta_I^2 }= {\left(x^{\mathcal{N}}\right)^2 \over \zeta_{\mathcal{N}}^2}  \, , 
\end{align}
which combined with Eq.~\eqref{eq:sum_zetak}  implies that 
\beq
\sum_{I=1}^{\mathcal{N}}  {(x^I)^4 \over \zeta_I^2 } = \frac{1}{\sum_{I=1}^{\mathcal{N}}  \zeta_{I}^2  }.
\eeq
We therefore find the potential on the attractor solution to be
\beq \label{eq:potential_attractor}
V_{\rm {at}}=  { r^4 \over (1+ r^2)^2 \sum_{I=1}^{\mathcal{N}}  \zeta_{I}^2} 
\eeq
(see Figure~\ref{fig:van_cross_coupl_equal_couplings}).
In terms of the original fields we find that they follow
\beq \label{eq:attractor_expression_ij}
{\phi^I \over \phi^J } =  \sqrt{\xi_J \over \xi_I}  {\zeta_I \over \zeta_J} \, .
\eeq

To demonstrate that this is a stable late-time solution we  calculate the second derivatives of the potential. The normalized masses (normalized with respect to $H$) are
\beq
\mathcal{M}_{IJ} \equiv {V_{,IJ} \over H^2} = 8(3 - \epsilon) \left( \delta_{IJ} + {\zeta_I \zeta_J \over \zeta_{\mathcal{N}}^2}  \right)  \, ,
\eeq
which are heavy, therefore the orthogonal fields are strongly stabilized. On the contrary, the solutions with at least one $x^{K}=0$ are all unstable as can be checked by direct substitution.

The number of $e$-folds generated on the attractor solution as a function of the initial radius is given by
\beq
N \approx - \int {V \over h^{rr} V_{,r}} d r    \, , 
\eeq
thus we need to calculate $h^{rr}$, i.e.~the component of the inverse metric in the $r,x$ coordinate system. The inverse component $h^{rr}$ is easily found when all $\xi_{I}$'s are of the same order because the $h$ metric becomes almost block-diagonal, or when there is one dominant combination of the non-minimal and self-coupling constant $\zeta$ which without loss of generality we can consider it to be $\zeta_{\mathcal{N}} \gg1$. For the latter case Eq.~\eqref{eq:x_k} shows that $x_{a}\ll 1$, hence the metric on the attractor solution becomes almost diagonal and we recover the single-field result \cite{Kaiser:2013sna,Greenwood:2012aj}
\beq \label{eq:nefolds}
N = {3 \over 4} r_0^2 +  \mathcal{O}({\xi_{\mathcal{N}}}^{-1} )\, .
\eeq
We have numerically checked  the validity of the previous equation at strong coupling. We expect deviations from the previous simple formula only when $\xi_I \approx 1$. Indeed, for two fields, parameterizing the $\xi$'s and $\zeta$'s as $\xi_1 = b \xi_2$, $\xi_2 = \xi$ and $\zeta_1 = c \zeta_2$ respectively the full expression is
\beq
N = \left(1 + r_0^2 \right) \left({3 \over 4} +  {1 + b^2 \over 8 \xi(1 + b^2 c)}  \right) - {3 \over 4} \ln (1 + r_0^2) \, ,
\eeq
which shows that the large $\xi$ limit is sufficient to derive the simple expression of Eq.~\eqref{eq:nefolds}.

We do not consider the limit of small non-minimal couplings,  $\xi_I \ll1$, where the effect of the geometry diminishes (at least for the last 55-60 $e$-folds) and the potential becomes simply  a sum of fields with quartic couplings.

\subsection{Non-zero cross couplings}
Let us consider a Jordan frame potential with cross-couplings of the form $g_{IJ} (\phi^I)^2 (\phi^J)^2$. Under the field redefinition used in Section~\ref{sec_vanishing_cc} the potential in the Einstein frame becomes 
\beq
V(r,x^I) =  {    r^4 \over (1+ r^2)^2}  \sum_{I,J=1}^{\mathcal{N}}   { g_{IJ}(x^I)^2 (x^J)^2 \over \xi_I \xi_J } =  {    r^4 \over (1+ r^2)^2}  \sum_{I,J=1}^{\mathcal{N}}   \Lambda_{IJ}(x^I)^2 (x^J)^2  \, ,
\eeq
where  we absorbed the non-minimal couplings into the self-couplings via $\Lambda_{IJ} \equiv g_{IJ}/(\xi_I \xi_J )$. 
We require the potential to be non-negative for all field values, so it is sufficient to consider either positive couplings ($\Lambda_{IJ}>0$) or positive eigenvalues of the matrix $\Lambda_{IJ}$. To find the minima we first rewrite the non-radial part as 
\beq
 \sum_{I,J=1}^{\mathcal{N}}   \Lambda_{IJ}(x^I)^2 (x^J)^2  =  \sum_{I,J=1}^{\mathcal{N}-1}   \Lambda_{IJ}(x^I)^2 (x^J)^2  + 2 \sum_{J=1}^{\mathcal{N}-1}  \Lambda_{\mathcal{N}J}(x^\mathcal{N})^2 (x^J)^2  + \Lambda_{\mathcal{N}\mathcal{N}}(x^\mathcal{N})^4 \, . 
\eeq
The minima are found as solutions to the system of algebraic equations 
\beq
4 \sum_{J=1}^{\mathcal{N}-1}  \Lambda_{KJ} (x^J)^2 x^K + 4 \Lambda_{\mathcal{N}K}(x^\mathcal{N})^2 x^K  -  4 \sum_{J=1}^{\mathcal{N}-1}  \Lambda_{\mathcal{N}J} (x^J)^2 x^K - 4  \Lambda_{\mathcal{N}\mathcal{N}}(x^\mathcal{N})^2 x^K = 0 \, .
\eeq
Excluding $x^K=0$, the minima are given by
\begin{align}
\label{eq:fkn}
\sum_{J=1}^{\mathcal{N}-1}  \left( \Lambda_{KJ} -\Lambda_{J \mathcal{N}} +\Lambda_{\mathcal{N}\mathcal{N}} - \Lambda_{K \mathcal{N}} \right) (x^J)^2 =  \Lambda_{\mathcal{N}\mathcal{N}} - \Lambda_{K \mathcal{N}} \, ,
\end{align}
or in a compact form
\beq
\sum_{J=1}^{\mathcal{N}}  \left( \Lambda_{KJ} -\Lambda_{\mathcal{N}J} \right) (x^J)^2 = 0 \, .
\eeq
Multiplying by the inverse matrix of $\Lambda$ yields
\beq
(x^I)^2= \sum_{K=1}^{\mathcal{N}} (\Lambda^{-1})_{KI}   \sum_{J=1}^{\mathcal{N}}\Lambda_{\mathcal{N}J} (x^J)^2  \Rightarrow  \sum_{K=1}^{\mathcal{N}}  (\Lambda^{-1})_{KI}    \sum_{J=1}^{\mathcal{N}}\Lambda_{\mathcal{N}J} (x^J)^2  = 1\, .
\eeq
Therefore, we find the non-trivial solution for each $x^I$ as
\beq \label{eq:xi2}
(x^I)^2 = { \sum_{K=1}^{\mathcal{N}}  (\Lambda^{-1})_{KI}  \over \sum_{I,K=1}^{\mathcal{N}}  (\Lambda^{-1})_{KI} } \, ,
\eeq
and the potential along the attractor as
\beq \label{eq:potential_2}
V_{\rm {at}} = { r^4 \over ( 1 + r^2 )^2}  { 1 \over  \sum_{I,K=1}^{\mathcal{N}}  (\Lambda^{-1})_{KI} }  \, .
\eeq 
We conclude that the non-trivial solution is fully determined by the inverse matrix of the redefined  couplings $\Lambda_{IJ}$. Note that Equations~\eqref{eq:xi2} and \eqref{eq:potential_2} collapse to Equations~\eqref{eq:zk} and \eqref{eq:potential_attractor}, respectively, with $\Lambda^{-1}_{II} = \zeta_I^2$, as expected. To study the stability of the various solutions we need the matrix of effective masses
\beq
\mathcal{M}_{KL} = {1 \over 3 H^2} \left(12 \sum_{J=1}^{\mathcal{N}}  \left( \Lambda_{KJ} -\Lambda_{J \mathcal{N}} \right) (x^J)^2 \delta^K_{~L} + 24  \left( \Lambda_{KL} -\Lambda_{L \mathcal{N}} \right) x^L x^K \right) \, .
\eeq
For inflation along a field axis 
$\phi^I$, which we take to be $\phi^{\mathcal{N}}$ without loss of generality,  we obtain a diagonal effective $\mathcal{M}$
\beq
\mathcal{M}_{KL} = {12 \over \Lambda_{\mathcal{N}\mathcal{N}}} \left( \Lambda_{K\mathcal{N}} -\Lambda_{\mathcal{N}\mathcal{N}} \right) \delta_{KL} \, ,
\eeq
and the solution becomes stable when $\Lambda_{K\mathcal{N}} > \Lambda_{\mathcal{N}\mathcal{N}}$. For the non-trivial solution we obtain
\beq
\mathcal{M}_{KL} =  { 24 \left( \Lambda_{KL} -\Lambda_{ L \mathcal{N}} \right) \over  \left( \sum_{I,J =1}^{\mathcal{N}}  (\Lambda^{-1})_{IJ} \right)^2 } \sum_{I,J}^{\mathcal{N}}  (\Lambda^{-1})_{IK}  (\Lambda^{-1})_{LJ}  \, ,
\eeq
and it will be stable if all eigenvalues have a positive real part. 

In general, radial evolution along some direction different than the original field axes is possible under certain conditions. For two fields we find the $x^I$ values as
\beq \label{eq:attractor_expression_fij}
\left(x^1 \right)^2 =  { \Lambda_{2 2} - \Lambda_{12}\over \Lambda_{11} -2 \Lambda_{12} + \Lambda_{22} } \, , \qquad \left(x^2 \right)^2 =  { \Lambda_{11} - \Lambda_{12}\over \Lambda_{11} -2 \Lambda_{12} + \Lambda_{22} } \, , 
\eeq
which exist under the conditions
\beq \label{eq:inequalities}
\Lambda_{12} < \Lambda_{11},\Lambda_{22} \,  \qquad \text{or}  \qquad \Lambda_{12} > \Lambda_{11},\Lambda_{22} \, .
\eeq
Moreover, the value of the potential along the attractor becomes
\beq
V_{\rm {at}} = { r^4 \over ( 1 + r^2 )^2}  {\Lambda_{11} \Lambda_{22} -\Lambda_{12} ^2 \over  \Lambda_{11} - 2\Lambda_{12} + \Lambda_{22} }  \, .
\eeq

\begin{figure}
\includegraphics[width=.45\textwidth]{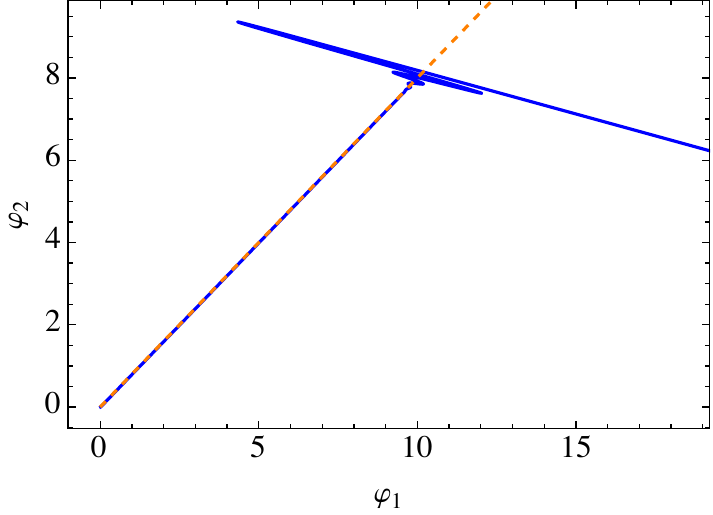}
\includegraphics[width=.45\textwidth]{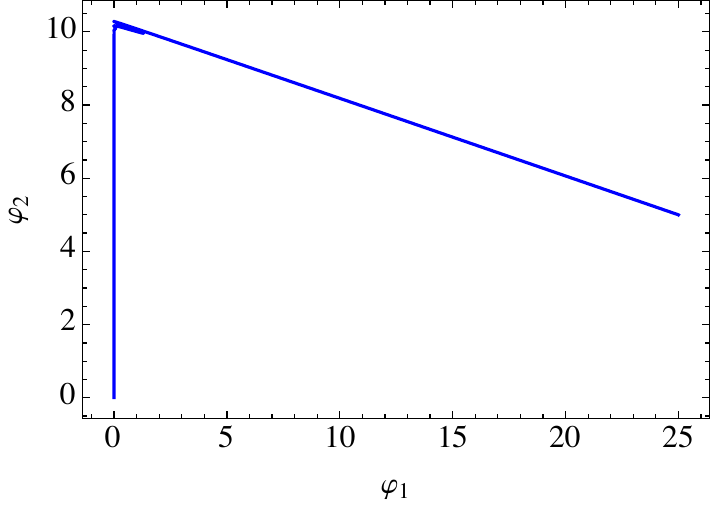}
\caption{Numerical (blue) versus the attractor expressions (orange, dashed) for a two-field realization with $\lambda_1=1,~\lambda_2=10,~\xi_1=2,~\xi_2=10$ and $\lambda_{12}=3$: The left plot matches the attractor expression \eqref{eq:attractor_expression_fij} whereas for the second set of parameters we find $\Lambda_{11}-\Lambda_{12}>0$,  $\Lambda_{22}- \Lambda_{12}<0$ and we conclude that the $\phi_1=0$ is the stable solution.}
\label{fig:cross_coupl}
\end{figure}

If the previous inequalities are not satisfied then only inflation along (some of) the field axes $\phi^I$ is possible. This is indeed the case for Figure~(2) of Refs.~\cite{Kaiser:2013sna,DeCross:2015uza} where the former inequalities are satisfied only for the green and black curves describing radial evolution for some $\theta \neq 0,\pi/2$ in the $\phi-\chi$ plane, whereas the red and blue curves display evolution along $\theta = 0$ and $\theta =\pi/2$ respectively, i.e.~the $\phi$ and $\chi$ axes. The effective masses for the non-trivial solution and the solutions $x^1,x^2=0$, i.e.~inflation along the $\phi^2$ and $\phi^1$ axes respectively are
\begin{align}
\mathcal{M}|_{\theta \neq 0} &=  24 {\left( \Lambda_{22} - \Lambda_{12} \right) \left( \Lambda_{11} - 2\Lambda_{12} + \Lambda_{22}\right)  \over \Lambda_{11} \Lambda_{22} -\Lambda_{12} ^2} \, , \\
\mathcal{M}|_{\phi^1=0} &=  12 { \Lambda_{12} - \Lambda_{22} \over  \Lambda_{22}} \, , \\
\mathcal{M} |_{\phi^2=0} &=  12 { \Lambda_{12} - \Lambda_{11} \over \Lambda_{11}} \, .
\end{align}
The sign of the previous three effective squared masses depends on the sign of $\lambda_{12}$. As we explained before if $\lambda_{12}<0$ the potential is non-negative when 
\beq
\Lambda_{11} \Lambda_{22} >\Lambda_{12}^2 \, , 
\eeq 
which combined with the requirement $\Lambda_{II}>0$ makes the first mass positive and the other two negative. On the contrary, when $\Lambda_{12}>0$ and only one of the two inequalities \eqref{eq:inequalities} is satisfied, i.e.~$\Lambda_{12} < \Lambda_{11}$ and $\Lambda_{12} > \Lambda_{22}$ or $\Lambda_{12} >\Lambda_{11}$ and $\Lambda_{12} < \Lambda_{22}$, the non-trivial solution does not exist and only one of the two field axes is stable. Finally, if all three solutions exist then we have the following two cases
\begin{align}
\Lambda_{12} < \Lambda_{11},~\Lambda_{22} \qquad  \Rightarrow  \qquad  \mathcal{M}|_{\theta \neq 0} > 0\, , \qquad  \mathcal{M} |_{\phi^1=0} \, , \mathcal{M} |_{\phi^2=0} <0 \,, \\
\Lambda_{12} > \Lambda_{11}, ~ \Lambda_{22} \qquad  \Rightarrow \qquad  \mathcal{M}|_{\theta \neq 0} < 0 \, , \qquad \mathcal{M} |_{\phi^1=0}\, , \mathcal{M} |_{\phi^2=0} > 0 \,.
\end{align}
Figure \ref{fig:cross_coupl} shows the agreement between our analytical results and the numerical evolution. 

Before concluding this section, we briefly comment on the case of non-zero mass terms $m_{I}^2 (\phi^I)^2$ in the potential. When adding these terms, the potential no longer admits a product-separable form
\beq
V(r,x^I) =  { r^2 \over (1+ r^2)^2}  \left( r^2 \sum_{I,J=1}^{\mathcal{N}}   \Lambda_{IJ}(x^I)^2 (x^J)^2 +  \sum_{I=1}^{\mathcal{N}}   { m_{I}^2 (x^I)^2\over \xi_I } \right)  \, ,
\eeq
while several new local minima emerge that behave effectively as a cosmological constant. This means that depending on the initial conditions the system can end in different places with drastically different inflationary evolution, and if it falls into the basin of attraction of one these minima, inflation will never end. However, while this is a mathematical fact, it will be realized physically only if (some of) the masses are comparable in size to the quartic terms. 
While simulating this potential can be easily accommodated within our code, the parameter space becomes too large and may obstruct the physical intuition we attempt to build in our current work. An extensive parameter scan would provide the fraction of parameter space that admits smoothly ending inflationary solutions. Although we expect inflation to proceed ``normally" apart from rather special ranges of mass values, it is an interesting problem, which
we leave for future work.

\section{Inverse distribution}
\label{app:inversedist}

In Section~\ref{sec:simulations} we provided numerically and analytically derived distributions for the number of $e$-folds and the Hubble parameter. The properties of the distribution of the number of $e$-folds (mean and variance) are straightforward to compute from the PDF's of the non-minimal couplings and the fields' amplitude, irrespective of the (Jordan frame) potential parameters.

The distribution of the Hubble scale along the attractor is more complicated. It does not depend on the field amplitude (as expected) and instead depends on the potential parameters and non-minimal couplings. More importantly, the denominator of Eq.~\eqref{eq:hubbleattractor} can be modeled (for large number of fields) as a Gaussian random variable, albeit taking only positive values, meaning that we expect its mean to be (much) larger than its standard deviation. If a random variable $y$ follows a normal distribution with mean $\mu_y$ and standard deviation $\sigma_y$, the random variable $z=1/y$ follows the reciprocal normal PDF with $f(z) = {1  \over \sqrt{2\pi} \sigma_y z^2} e^{-{1\over 2}\left ( {{1\over z}-\mu_y} \right )^2/ \sigma_y^2}$. For our purposes, we can safely assume that the variable $y$ does not take negative values, meaning that $\mu_y\gg \sigma_y$. We can thus calculate the moments of $z$ by integrating $z\in (0,\infty)$ or by extending the regime of integration $z\in (-\infty,\infty)$, since the ``portion" of the PDF for $z<0$ is exponentially suppressed. 
\beq
\E[z]=\int_{-\infty}^\infty z f(z) dz
= \int_{-\infty}^\infty {1\over x} {1\over \sqrt{2\pi}\sigma_y} e^{-{1\over 2}\left ( {{x}-\mu_y} \right )^2/ \sigma_y^2} \ud x
\eeq
where we performed a change of variables $x=1/y$. The integrand is strongly centered around $\mu_y$ and thus we can use the Taylor expansion of $x$ around $x=\mu_y$ and compute the integral as an infinite sum. Interestingly for $\mu_y\sigma_y\gg 1$ we need only keep the leading order term  $\mu_z \equiv \E[z] \simeq \mu_y^{-1}$. The same can be done for the variance, by computing $\sigma_z^2= \E[z^2]- \E[z]^2$. However, keeping only the leading order terms for $\E[z^2],\E[z]^2$ makes the variance vanish. We this need to go the next non-trivial order in both terms, arriving at $\sigma_z\simeq \sigma_y/\mu_y^2$. Finally, in the regime of validity of these approximations, the reciprocal normal distribution is practically indistinguishable from a normal distribution
\beq
f(z) = {1  \over \sqrt{2\pi} \sigma_y z^2} e^{-{1\over 2}\left ( {{1\over z}-\mu_y} \right )^2/ \sigma_y^2} \simeq
{1  \over \sqrt{2\pi} \sigma_z} e^{-{1\over 2}\left ( {z-\mu_z} \right )^2/ \sigma_z^2}
\simeq
{\mu_y^2  \over \sqrt{2\pi} \sigma_y} e^{-{\mu_y^4\over 2\sigma_y^2}\left ( {z-{1\over \mu_y}} \right )^2} \, .
\eeq
This allows for a very simple analysis of the distribution of Hubble parameters based on the distribution of non-minimal couplings $\xi_I$ and the fields' self-couplings $\lambda_I$.

\bibliography{NMCrefs}{}

\end{document}